%% file: OJCOMS-02583-2024.tex
\documentclass{IEEEoj}

\usepackage[active]{srcltx} 
\usepackage[cmex10]{amsmath}
\usepackage{amsfonts}
\usepackage[final]{graphics}
\usepackage[final]{graphicx}
\usepackage{amsbsy}
\usepackage{amsbsy}
\usepackage{amssymb}
\usepackage{url}
\usepackage{enumerate}
\usepackage{cite}
\usepackage{psfrag}
\usepackage{color}
\usepackage{euscript}
\usepackage[utf8]{inputenc}
\usepackage{algorithmic}
\usepackage[ruled,vlined]{algorithm2e}
\usepackage{soul}
\usepackage{multirow}

\def\BibTeX{{\rm B\kern-.05em{\sc i\kern-.025em b}\kern-.08em
    T\kern-.1667em\lower.7ex\hbox{E}\kern-.125emX}}
\AtBeginDocument{\definecolor{ojcolor}{cmyk}{0.93,0.59,0.15,0.02}}
\def\OJlogo{\vspace{-4pt}\hskip-4pt\includegraphics[height=18pt]{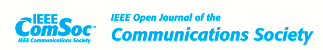}}

\newcommand{\bm}[1]{{\mathbf{#1}}}
\newcommand{\Es}{{\mathbb{E}}}          
\newcommand{\rank}{{\text{rank}}}
\newcommand{\diag}{{\text{diag}}}
\newcommand{\trace}{{\text{tr}}}

\newcommand{\I}{\bm{I}}
\newcommand{\Jb}{\bm{J}}
\newcommand{\Zero}{\bm{O}}

\newcommand{\Lcp}{L_{\text{cp}}}
\newcommand{\Lf}{L_{\text{filter}}}
\newcommand{\y}{y}

\newcommand{\xa}{x_{\text{A}}}

\newcommand{\ba}{s_{\text{A}}}
\newcommand{\bs}{s_{\text{T}}}
\newcommand{\yb}{\bm y}
\newcommand{\ybtilde}{\widetilde{\bm y}}
\newcommand{\Yb}{\bm Y}
\newcommand{\Db}{\bm D}
\newcommand{\Eb}{\bm E}
\newcommand{\Fb}{\bm F}

\newcommand{\Mb}{\bm M}
\newcommand{\Mbtilde}{\widetilde{\bm M}}
\newcommand{\hb}{\bm h}

\newcommand{\ab}{\bm a}

\newcommand{\wb}{\bm w}
\newcommand{\dw}{\bm d}
\newcommand{\dwtilde}{\widetilde{\bm d}}
\newcommand{\Ab}{\bm A}
\newcommand{\gb}{\bm g}
\newcommand{\Rb}{\bm R}
\newcommand{\vb}{\bm v}
\newcommand{\Bb}{\bm B}
\newcommand{\Lb}{\bm L}

\newcommand{\babpsk}{b_{\text{A}}}
\newcommand{\bab}{\bm s_{\text{A}}}
\newcommand{\bsb}{\bm s_{\text{T}}}
\newcommand{\Hatildezero}{\overline{\bm H}_{\text{A},j}^{(0)}}
\newcommand{\Hatildeuno}{\overline{\bm H}_{\text{A},j}^{(1)}}
\newcommand{\Hstildezero}{\overline{\bm H}_{\text{T},j}^{(0)}}
\newcommand{\Hstildeuno}{\overline{\bm H}_{\text{T},j}^{(1)}}

\newcommand{\xs}{x_{\text{T}}}
\newcommand{\xsk}{x_{\text{T},k}}

\newcommand{\Ks}{K_{\text{T}}}
\newcommand{\Cset}{\mathbb{C}}
\newcommand{\Rset}{\mathbb{R}}

\newcommand{\Zset}{\mathbb{Z}}

\newcommand{\eqdef}{\triangleq}

\newcommand{\herm}{\text{H}}
\newcommand{\trasp}{\text{T}}

\newcommand{\pot}{\EuScript{P}}

\def\bdm#1\edm{\begin{displaymath}#1\end{displaymath}}
\def\be#1\ee{\begin{equation}#1\end{equation}}
\def\barr#1\earr{\begin{align}#1\end{align}}

\newcommand{\IeeeTIT}{{\em IEEE Trans.\ Inf. Theory\/}}
\newcommand{\IeeeTSP}{{\em IEEE Trans.\ Signal Process.\/}}
\newcommand{\IeeeTCOMM}{{\em IEEE Trans.\ Commun.\/}}
\newcommand{\IeeeCOMMLETT}{{\em IEEE Commun.\ Lett.\/}}

\newcommand{\IeeeWCOMMLETT}{{\em IEEE Wireless Commun.\ Lett.\/}}
\newcommand{\IeeeTWC}{{\em IEEE Trans.\ Wireless Commun.\/}}

\newcommand{\IeeeTVT}{{\em IEEE Trans.\ Veh. Technol.\/}}

\newcommand{\IeeeJSTSP}{{\em IEEE J.\ Select.\ Topics Signal Process.\/}}

\newcommand{\IeeeCOMMMAG}{{\em IEEE Commun.\ Magazine\/}}

\input{arXiv.tex}

\begin{document}

\receiveddate{4 July, 2024}
\reviseddate{5 August, 2024}
\accepteddate{26 August, 2024}
\publisheddate{Day Month, 2024}
\currentdate{26 August, 2024}
\doiinfo{xxx}

\title{Channel State Acquisition in Uplink NOMA for
Cellular-Connected UAV: Exploitation of Doppler
and Modulation Diversities
}

\author{DONATELLA~DARSENA\authorrefmark{1} (SENIOR MEMBER, IEEE), 
IVAN~IUDICE\authorrefmark{2}, AND 
FRANCESCO~VERDE\authorrefmark{1} (SENIOR MEMBER, IEEE)}
\affil{Department of Electrical Engineering and
Information Technology, University Federico II, Naples I-80125, Italy}
\affil{Reliability \& Security Department, Italian Aerospace Research Centre (CIRA),
Capua I-81043, Italy}
\corresp{CORRESPONDING AUTHOR: F.~VERDE (e-mail: f.verde@unina.it).}
\authornote{This work was partially supported by the European Union under the Italian National Recovery and Resilience Plan (NRRP) of NextGenerationEU, partnership on ``Telecommunications of the Future" (PE00000001 - program ``RESTART")
and the CIRA project ``MATIM".}
\markboth{Channel State Acquisition in Uplink NOMA for
Cellular-Connected UAV}{Darsena \textit{et al.}}

\begin{abstract}

Integration of unmanned aerial vehicles (UAVs) 
for surveillance or monitoring applications
into fifth generation (5G) New Radio (NR)  
cellular networks is an intriguing problem that has 
recently tackled a lot of interest in both academia and industry. 
For an efficient spectrum usage, we consider a recently-proposed 
sky-ground nonorthogonal multiple access (NOMA) scheme, where
a cellular-connected UAV acting as  aerial user (AU) and a static terrestrial user (TU) 
are paired to simultaneously transmit their uplink signals to a base station (BS) 
in the same time-frequency resource blocks. 
In such a case, due to the highly dynamic nature of the UAV, 
the signal transmitted by the AU experiences both time dispersion due to multipath 
propagation effects and frequency dispersion caused by Doppler shifts. 
On the other hand, for a static ground network, 
frequency dispersion of the signal transmitted by the TU is negligible  
and only multipath effects have to be taken into account. 
To decode the superposed signals at the BS through 
successive interference cancellation,  accurate estimates
of both the AU and TU channels are needed. In this paper, we 
propose channel estimation procedures
that suitably exploit the different circular/noncircular modulation 
formats ({\em modulation diversity}) and 
the different almost-cyclostationarity features ({\em Doppler diversity}) 
of the AU and TU by means of widely-linear time-varying processing. 
Our estimation approach is semi-blind since 
Doppler shifts and time delays of the AU are estimated  based on 
the received data only,
whereas the remaining relevant parameters of 
the AU and TU channels are acquired relying also on the available training symbols,
which are transmitted by the AU and TU in a nonorthogonal manner.
Monte Carlo numerical results demonstrate that 
the proposed channel estimation algorithms can 
satisfactorily acquire all the relevant parameters
in different operative conditions.

\end{abstract}

\begin{IEEEkeywords}
Almost-cyclostationarity, channel estimation, circular and noncircular modulations,
doubly selective channels, Doppler diversity, 
5G New Radio (NR), modulation diversity, 
non-orthogonal multiple access (NOMA), unmanned aerial vehicles (UAVs), uplink.
\end{IEEEkeywords}

\maketitle
\acceptednotice

\section{Introduction}

\IEEEPARstart{W}{ith} the development of unmanned aerial vehicles (UAVs), computer vision, 
and sensor technology, UAV systems have been increasingly employing 
in civilian and commercial applications, such as surveillance and monitoring, due to
their ability to quickly cover large and difficult-to-reach areas.
The recently-introduced concept of cellular-connected UAVs 
\cite{Zeng.2019-1,Zeng.2019-2,Mei.2021} represents a viable strategy to
integrate UAVs in fifth generation (5G) New Radio (NR) networks, by allowing them 
to transmit data to the cellular  network as aerial users (AUs). 
3GPP Release 18 is introducing 5G NR support to enhance 
the safe use of UAVs for commercial and leisure applications.
Nevertheless, the integration of UAVs into cellular networks introduces two 
interelated basic challenges \cite{Lin,New}.
First, existing 5G cellular infrastructures are designed to serve terrestrial users (TUs)
and, thus, the antennas of the base stations (BSs) look downward: such
a tilting might cause AUs to be served by the side lobes of the BSs and, consequently, 
they suffer from a reduction in antenna gain. 
Second, due to their mobility capabilities, AUs have more versatile movements than 
TUs, thereby introducing Doppler (frequency) shifts, which result in
carrier frequency offset, inter-carrier interference, and reduced
channel coherence time.

\IEEEpubidadjcol

The feasibility of using existing cellular networks
to support UAVs in the low-altitude airspace has been
demonstrated in \cite{Nokia, Huawei,Amer.2019}. 
Specifically, it has been shown that
the favorable line-of-sight (LoS)
propagation conditions
for UAVs flying in the sky can 
compensate for 
the reduced gain of antenna 
side lobes of the BSs, provided that 
UAVs fly below $120-200$ m.
A first consequence of flying at lower altitudes
is that the AU-to-BS channel exhibits multipath 
components (MPCs) consisting of an LoS link 
with high probability and a cluster of reflected, 
delayed paths \cite{Haas}. 
Severe Doppler shifts are caused
by the high carrier frequency and high velocity of the UAV,
and they are influenced by the angular
distribution of the scattered components.
Indeed, different
MPCs may have largely different Doppler shifts.

To support AUs in cellular networks, there are two possible 
multiple access schemes. In the case of 
orthogonal multiple access (OMA) 
\cite{VanDerBergh, Afonso, Ngu, Xue, Azari, Chandhar, Geraci, Amer, Challita-1, Challita-2}, 
AUs and TUs are served using orthogonal resource blocks (RBs). 
On the other hand, {\em non-orthogonal
multiple access (NOMA)} allows AUs and TUs to simultaneously share the same spectrum through 
power-domain or code-domain multiplexing: in this case,  
successive interference cancellation (SIC) is used at the BS 
to separate the signals. 
In many cases, NOMA ensures superior spectral efficiency with respect to OMA
\cite{Ding.2017, Shin, Islam, Sari, Ding2014,Wei.2020}.

\subsection{NOMA for cellular-connected UAVs}

Outage probabilities in downlink and uplink NOMA transmissions have been evaluated in 
\cite{Zaidi} based on instantaneous distinct signal power for devices
including UAVs.
In \cite{New-2}, an aerial-ground NOMA scheme that pairs the AU and TU for data and control links
has been investigated, by exploiting the asymmetric features of the channels and rate demands of the AU and TU in the downlink communication.
The authors in \cite{Mei} have proposed a cooperative downlink NOMA scheme to cancel
co-channel interference via SIC in ideal two-cell networks. 
A cooperative NOMA scheme that exploits existing backhaul links among BSs
for SIC operations has been proposed in \cite{Mei-2}.
An uplink NOMA has been proposed in \cite{Mu} to serve AU and TU,
where AU trajectory and its cell-association order are jointly optimized,
whereas  the use of uplink NOMA for AUs and TUs has been studied 
via stochastic geometry, where the mobility of the AUs is taken into account. 
The probability that the achievable data rate of both the AU and TU 
exceeds the respective target rates is calculated in \cite{Sena},
where the minimum
height that the AU needs to fly, at each transmission point
along the given trajectory, is also numerically determined
in order to satisfy a certain quality
of service constraint.
Finally, the authors in \cite{Pang} have derived  the optimal 
precoding that maximizes the uplink sum-rate of an AU and multiple TUs. 

{\em A common underlying approach of all the aforementioned works 
\cite{Zaidi,New-2,Mei,Mei-2,Mu,Sena,Pang} is that 
Doppler shifts due to the UAV motions are  a cause of 
performance degradation rather than a source of diversity}.
Indeed, it is tacitly assumed that large Doppler shifts
have been previously compensated for or, as in \cite{Sena},
that the AU transmits its data to the BS by maintaining a fixed 
position over selected points along a given trajectory
(i.e., transmission during hovering flight).
However, compensation of Doppler shifts is
a challenging task in multipath channels, except 
for the case  when the MPCs of the AU-to-BS channel 
have a very similar Doppler shift, which is a situation 
that may not be fulfilled in the low-altitude airspace \cite{Haas}. 
Moreover, allowing the UAV to transmit only in the 
hovering-flight state along its trajectory entails a waste of both 
communication and power resources.
{\em Another shortcoming of the aforementioned papers \cite{Zaidi,New-2,Mei,Mei-2,Mu,Sena,Pang}
is that channel estimation issues have not been accounted at all}.
In a NOMA setting, channel acquisition of both the AU and TU channels is crucial 
to obtain accurate SIC at the BS.

\subsection{Modulation formats for double-selective channels}
The choice of the digital modulation format crucially determines
accuracy and signaling overhead of channel estimation. 
{\em Orthogonal time frequency space (OTFS)} and
{\em orthogonal frequency division multiplexing (OFDM)} are
digital modulation formats designed to handle doubly 
(i.e., time-and-frequency) selective channels.
OFDM modulation directly places symbols in the 
time-frequency (TF) domain, whereas 
OTFS is a 
modu\-lation technique in which symbols are 
transmitted in the  delay-Doppler (DD) domain \cite{Hadani.2017}.

OFDM is immune to the frequency-selective nature
of the wireless channel. However, if frequency dispersion is also introduced by
the channel  in the form of Doppler shifts, the orthogonality of
the subcarriers is destroyed, thus resulting in intercarrier interference (ICI).
On the other hand, OTFS might cope with multipath fading
and Doppler shifts more effectively, but such a robustness comes at the cost 
of a nonnegligible increase in signal processing complexity.

OTFS modulation has been used
in \cite{Zing.2019} for the implementation of NOMA among users  
with different mobility profiles: specifically, a high-mobility user
is allowed to share the spectrum with multiple low-mobility NOMA users.
Although the specific case of an AU is not explicitly considered in 
\cite{Zing.2019,Zing.2020}, their framework may be 
adapted to the scenario of cellular-connected UAVs, by regarding 
the high-mobility user as an UAV.
However,  perfect channel know\-ledge is assumed in \cite{Zing.2019,Zing.2020}. 

The accuracy and signaling overhead of channel estimation techniques for OTFS modulation strictly
depend on the values assumed by the delays and Doppler shifts.
Indeed, in the case of integer delays and Doppler shifts, 
the received symbol corresponding to a given channel 
path is well-localized in the corresponding DD bin \cite{Zing.2019,Zing.2020,Tho.2022}. 
On the other hand, for fractional delays and Doppler shifts,
each transmitted symbol spreads into adjacent bins, resulting
in received symbols interfering with each other \cite{Mup.2023}.
Such an inter-path interference (IPI) is a source of performance degradation for 
channel estimation performed directly in the DD domain. 
In the case of training-based channel identification,
which consists into placing pilot symbols in the DD grid,
one way to reduce the detrimental effects of IPI is to introduce 
guard symbols \cite{Rav.2019}. However, the guard
interval of the pilots needs to include the entire Doppler domain
within the range of the maximum delay, which results
in a large pilot overhead. To avoid waste of communication resources, cancellation procedures
of IPI might be carried out before channel estimation \cite{Mup.2023,Hu.2024},
which inevitably increase implementation complexity of OTFS coherent reception. 

Channel estimation for OFDM modulation is directly performed in the TF domain 
and, with respect to OTFS, it can face with fractional delays and Doppler shifts
more effectively. Moreover, OFDM was extensively used in many standards
of wireless systems, e.g.,  digital audio/video
broadcasting (DAB/DVB),  Wi-Fi local networks, worldwide interoperability for 
microwave access (WiMAX), long term evolution (LTE) and LTE-Advanced, 
and, it has been adopted in the 5G NR network. OFDM is still considered 
one of the potential candidates for beyond-5G communication systems. 
However, since OFDM is more sensitive to Doppler shifts than 
OTFS, a higher level of estimation accuracy is demanded.
Despite the fact that channel estimation for OFDM is a
well-investigated topic, to the best of our knowledge,
there are no effective channel estimation metho\-ds available yet 
for NOMA aerial and terrestrial OFDM uplink channels.

\subsection{Contribution and organization}

We consider a NOMA uplink scheme,
which is referred to  as {\em sky-ground (SG)} NOMA,  
where a flying AU is paired with a static (i.e., fixed) TU. 
The AU and TU have different propagation conditions 
and their received power can be rather different at the BS, 
regardless of their transmit power: indeed, the air-to-cellular (A2C)
channel between the AU and the elevated terrestrial BS
is  doubly selective with significant Doppler shifts,\footnote{The
air-to-cellular channel is different from the air-to-ground one due
to the non-negligible height of terrestrial BSs.}
whereas the TU-to-BS channel is 
predominately frequency-selective with 
negligible Doppler effects. 
We propose to {\em exploit} the different mobile profiles
of the AU and TU rather than to {\em counteract} the Doppler
effects induced by the mobility of the UAV. In particular, 
the AU is allowed to continuously transmit along all its trajectory.

To counteract the frequency-selectivity of both
channels, the AU and TU resort to 
the OFDM  transmission format.
To improve the user separation capability of the BS, 
we propose to exploit  a further source of diversity 
between the two users in the modulation domain: in 
both training (or pilot) and data phases, 
the TU transmits {\em circular} symbols, whereas the AU 
employs {\em noncircular} modulation.

Acquisition of channel state information (CSI) in uplink 
NOMA is a challenging task, since channel estimation 
in current wireless data communications standards 
is mainly based on training sequences sent 
from the users to the BS that are 
orthogonal in either time or frequency domain.
Orthogonal RBs for training adversely affect 
spectral efficiency and system performance, especially in the uplink. 
We consider a nonorthogonal training scheme where the training data of one 
user is contaminated by either the pilots or the information-bearing data
of the other user. In this challenging scenario, we propose 
a two-stage approach: first,  
the channel parameters of the AU 
are semi-blindly estimated by also exploiting the 
{\em almost-cyclostationarity (ACS) properties} 
 \cite{Gardner-book} of the received signal
that depend on the Doppler shifts of the AU-to-BS channel ({\em Doppler diversity}), as well as 
the different circular/noncircular features 
of the AU and TU ({\em modulation diversity});
subsequently, capitalizing on the CSI knowledge of
the AU, the overall channel of the TU is acquired with a 
pilot-aided improved widely-linear (WL) estimator \cite{Picinbono}. 
The performances of the proposed channel 
estimation algorithms are 
validated through extensive Monte Carlo numerical examples.

The paper is organized as follows.  The system model of the considered
uplink SG-NOMA  is described in Section~\ref{sec:system}.
The proposed estimation strategies for the AU and TU
are developed in Sections~\ref{sec:estimation-AU} and
\ref{sec:estimation-TU}, respectively.
Monte Carlo numerical results are reported in Section~\ref{sec:simul}
in terms of channel estimation accuracy.
Finally, the main results obtained 
in the paper are summarized in Section~\ref{sec:concl}.

\subsection{Basic notations}
Upper- and lower-case bold letters denote matrices and vectors;
the superscripts
$*$, $\trasp$, $\herm$, and $-1$
denote the conjugate,
the transpose, the Hermitian (conjugate transpose),
and the inverse of a matrix;
$\mathbb{C}$, $\mathbb{R}$ and $\mathbb{Z}$ are
the fields of complex, real and integer numbers;
$\mathbb{C}^{n}$ $[\mathbb{R}^{n}]$ denotes the
vector-space of all $n$-column vectors with complex
[real] coordinates;
similarly, $\mathbb{C}^{n \times m}$ $[\mathbb{R}^{n \times m}]$
denotes the vector-space of all the $n \times m$ matrices with
complex [real] elements;
$\delta(\tau)$ is the Dirac delta;
$\delta_n$ is the Kronecker delta, i.e., $\delta_n=1$ when $n=0$
and zero otherwise;
$\jmath \eqdef \sqrt{-1}$ denotes the imaginary unit;
$\text{max}(x,y)$  [$\text{min}(x,y)$] returns the maximum [minimum] 
between $x \in \Rset$ and $y \in \Rset$;
$\lfloor x \rfloor$
rounds $x \in \Rset$ to the nearest integer less than or equal to $x$;
the (linear) convolution operator is denoted with $\ast$;
$\frac{\partial}{\partial x*} f$ is the complex derivative 
of the real-valued scalar function $f$ with respect to $x^*$
\cite{Hj};
$\mathbf{0}_{n}$, $\Zero_{n \times m}$ and $\I_{n}$
denote the $n$-column zero vector, the $n \times m$
zero matrix and the $n \times n$ identity matrix; 
$\Jb_n$ is given by 
\be
\Jb_n  \eqdef \begin{pmatrix}
\bm{O}_{n \times n} & \I_{n} \\
\I_{n} & \bm{O}_{n \times n}
\end{pmatrix} \in \Rset^{(2 n) \times (2 n)} \:;
\label{eq:Jmat}
\ee
$\{\mathbf{a}\}_{\ell}$ is the $\ell$th entry of $\bm{a} \in \Cset^n$;
$\{\mathbf{A}\}_{\ell,\ell}$ and
$\mathrm{trace}(\mathbf{A})$ denote
the $\ell$th diagonal entry and the trace of $\mathbf{A} \in \Cset^{n \times n}$;
$\mathrm{rank}(\mathbf{A})$ is the rank of $\mathbf{A} \in \Cset^{n \times m}$;
$\otimes$ is the Kronecker product;
$\|\Ab\| \eqdef [\trace(\Ab\,\Ab^\herm)]^{1/2}$ denotes the 
Frobenius norm of $\Ab \in \Cset^{n \times m}$ \cite{Horn};
$\mathbf{A}= \diag (a_{0}, a_{1}, \ldots,
a_{n-1}) \in \mathbb{C}^{n \times n}$ is diagonal;
$\langle \cdot \rangle$ represents
infinite-time temporal averaging and $\Es[\cdot]$ denotes ensemble averaging;
$[\cdot]_2$ stands for modulo-$2$ operation;
finally, with denote with $x(t)$ and $x[n]$ continuos-time and 
discrete-time signals, respectively.

\subsection{Special functions and matrices}
\label{sec:pre}
The Dirichlet (or periodic sinc) function  is defined as 
\be
{\rm D}_n(x) \eqdef \frac{\sin(\pi x n)}{\sin(\pi x)} \, e^{-j \pi (n-1)x}
\label{eq:Dic}
\ee
for each $x \in \Rset$.
$\bm{W}_n \in \mathbb{C}^{n \times n}$ is the 
unitary symmetric $n$-point inverse discrete Fourier transform 
(IDFT) matrix, whose $(m+1,p+1)$-th entry is given by
$\frac{1}{\sqrt{n}} \, e^{\jmath \frac{2 \pi}{n} m p}$ for
$m,p \in \{0,1,\ldots, n-1\}$,  
and its 
inverse $\bm{W}_n^{-1}=\bm{W}_n^\herm$ is the $n$-point 
discrete Fourier transform (DFT) matrix;
$\Fb \in \Rset^{P \times P}$
and $\Bb \eqdef \Fb^\trasp \in \Rset^{P \times P}$ denote 
the {\em forward shift} 
and {\em backward shift} matrices, 
whose first column and the first row are given by
$[0, 1, 0, \ldots, 0]^\trasp$
and $[0, 1, 0, \ldots, 0]$, respectively.

\section{System model and basic assumptions}
\label{sec:system}

The considered single-cell network scenario  
encompasses an elevated BS, mobile AUs, and static TUs. 
The BS is located within the cell at a
fixed height $H_\text{BS}$,
whereas the AU flies at an altitude
$H_\text{A}$ above the ground.\footnote{The proposed 
estimation algorithms can be straightforwardly modified
to account for the case where the AU varies its height 
along the trajectory.}
We restrict ourselves to uplink communications since 
advanced signal detection algorithms of NOMA are more affordable
at the BS \cite{Wei.2020}.
A uniform linear array (ULA) is used
at the BS with $J$ antennas.
The AUs are represented by single-antenna UAVs that 
are employed for surveillance or monitoring a given area
within a cell and 
transmit data to the corresponding BS over given A2C links.
According to 5G NR channel access schemes for terrestrial
cellular networks \cite{NR.2019}, we assume that TUs transmit to the BS 
using the OMA scheme and, hence, there is no 
interference among the TUs in the cell.
The extension of the proposed approach to the case in which 
nonorthogonal RBs are assigned to TUs in the uplink is discussed
in Subsections~III-\ref{sec:NOMA-TU-A} and IV-\ref{sec:NOMA-TU-T}.

\begin{figure}[t]
\centering
\includegraphics[width=\columnwidth]{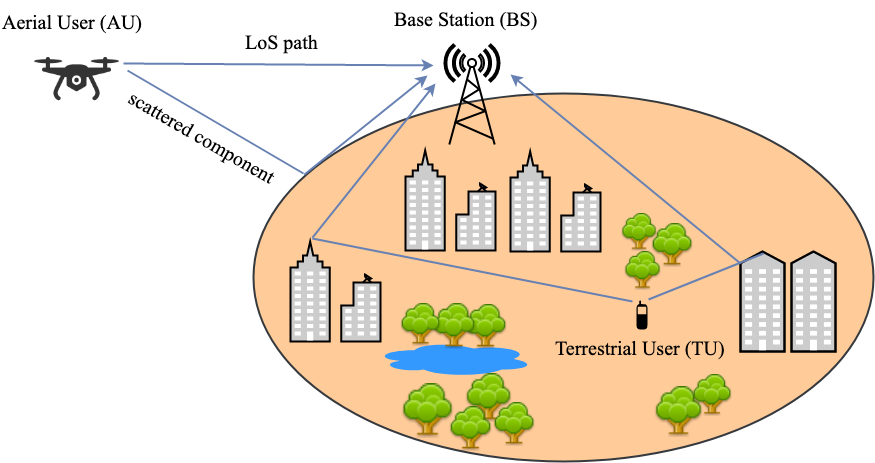}
\caption{Uplink sky-ground NOMA: the UAV
and the terrestrial user transmit their own uplink signals to
the BS in the same RBs.}
\label{fig:fig_1}
\end{figure}

In this paper, we consider an uplink SG-NOMA protocol, where 
single-antenna TUs share their RBs with the AUs. 
As in \cite{Sena}, we assume that each AU is randomly 
paired with a different static (i.e., fixed) TU 
(see Fig.~\ref{fig:fig_1}).
Hence, the scenario of interest boils down to the case  
of a single AU and a paired TU.\footnote{In principle,
multiple AUs can be 
paired with a single static TU. In this case, the proposed
estimation framework might be extended straightforwardly, provided that
the superimposed AU signals exhibit different ACS properties 
at the BS, i.e., the AU-to-BS channels are characterized by 
different Doppler shifts.
}
The two-user scenario represents the case of major practical 
interest for NOMA, since it results in acceptable 
interference levels \cite{Ding.2016}.
Moreover, we assume that inter-cell interference (ICI) coordination techniques are
employed to mitigate terrestrial ICI \cite{Kosta.2013, Hanza.2013}.
When the UAV flies in the low-altitude airspace \cite{Nokia, Huawei,Amer.2019},
the LoS probability between the UAV  and the neighboring co-channel BSs may be  
very low and ICI coordination (ICIC) only needs to involve a few BSs 
in a local region to exchange control information via high-speed backhaul links.
In this case, the use of more sophisticated aerial-ground interference
mitigation/cancellation procedures \cite{Mei.2021} might be avoided.
Additionally, since each AU uses NOMA with a paired TU in the same cell, terrestrial ICIC
indirectly limits aerial-ground interference between the UAV and 
co-channel terrestrial communications. 
Thus, in this work, we focus on the equivalent scenario 
with a single BS and a single AU-TU pairing in a given cell,
without ICI.

\subsection{Transmit signal model}

Both users 
employ OFDM transmission 
to cope with the time-dispersive 
nature of the channel.

The independent symbol streams 
emitted  by the AU and the TU ($n \in \mathbb{Z}$) are
denoted with $\{\ba[n]\}$ and $\{\bs[n]\}$, 
respectively.\footnote{Throughout the
paper, the subscripts A and T
refer to the AU (i.e, UAV) 
and the TU, respectively.}
We model $\{\ba[n]\}$ as a sequence of 
zero-mean unit-variance 
independent and identically distributed (i.i.d.) complex
{\em noncircular} random variables \cite{SS-book},
with second-order moment $\Es(\ba^2[n]) \neq 0$.
On the other hand, $\{\bs[n]\}$ is modeled as a sequence of 
zero-mean unit-variance i.i.d. complex
{\em circular} random variables, i.e., 
$\Es(\bs^2[n])=0$.
5G NR has been natively designed
for circular modulation schemes, such as
phase shift keying (PSK) and 
quadrature amplitude modulation (QAM).
Noncircular modulation schemes have been subsequently introduced 
for the uplink data channel (physical uplink
shared channel- PUSCH) and control channel (physical uplink
control channel - PUCCH) transmissions \cite{3GPP-1, 3GPP-2, 3GPP-3}, 
such as $\pi/2$ binary PSK (BPSK).
These waveforms, when
combined with an appropriate spectrum shaping, enable low
peak-to-average-power ratio transmissions without
compromising the error rate performance, which is
a desirable property for UAV communications.  
We will show herein that exploitation of the
different second-order statistics (i.e, 
noncircular versus circular) of the AU
and TU signals 
is instrumental into acquiring 
CSI in NOMA.  

Both users share the same $M$ orthogonal subcarriers.
Let $s_{\text{TX}}^{[m]}[\ell] \eqdef s_{\text{TX}}[\ell \, M + m]$
represent the symbol transmitted in the $\ell$th data block
on the $m$th subcarrier, with $\text{TX} \in \{\text{A},\text{T}\}$,
the vector
$\bm{s}_\text{TX}[n] \eqdef (s_\text{TX}^{[0]}[n], s_\text{TX}^{[1]}[n],\ldots,
s_\text{TX}^{[M-1]}[n])^\trasp \in \Cset^M$ is
subject to IDFT
and cyclic prefix (CP) insertion, thus yielding $\bm{u}_\text{TX}[n] \eqdef (u_\text{TX}^{[0]}[n], u_\text{TX}^{[1]}[n],\ldots,
u_\text{TX}^{[P-1]}[n])^\trasp= \bm{I}_\text{cp} \, \bm{W}_M \, \bm{s}_\text{TX}[n]$,
where $\bm{I}_\text{cp} \in \mathbb{R}^{P \times M}$ models 
the insertion of a CP of length $\Lcp$, with $P \eqdef M+\Lcp$, and $\bm{W}_M$ is the $M$-point
IDFT (see Subsection~I-D).
The precoded data vector $\bm{u}_\text{TX}[n]$ undergoes parallel-to-serial (P/S) conversion,
and the resulting data sequence $\{u_\text{TX}[\ell]\}$
feeds a digital-to-analog converter (DAC), operating at rate
$1/T_\text{c}=P/T_\text{s}$, where $u_\text{TX}[\ell \, P+q]=u_\text{TX}^{[q]}[\ell]$,
 for $q \in \{0,1.\ldots, P-1\}$,
the OFDM symbol period is $T_\text{s}$, whereas
$T_\text{c}$ is the sampling period.
The baseband transmitted continuous-time 
signal at the output of the DAC is given by
\be
x_\text{TX}(t) = \sum_{\ell=-\infty}^{+\infty} \sum_{q=0}^{P-1} u_{\text{TX}}^{[q]}[\ell]  \, 
\psi_\text{DAC}(t- q \, T_\text{c}-\ell \, T_\text{s}) 
\label{eq:xell}
\ee
for $\text{TX} \in \{\text{A}, \text{T}\}$, where $\psi_\text{DAC}(t)$
denotes the impulse response of the DAC.

\subsection{Channel model for terrestrial user}

The far-field channel between the TU and the BS is 
invariant over a coherence time 
$T_\text{coh} \eqdef N_\text{coh} \, T_\text{s}$, 
with $N_\text{coh} \gg 1$, wherein 
it can be modeled as a {\em linear time invariant (LTI)} system with impulse response:
\be
\hb_\text{T}(\tau)=  \sum_{k=1}^{\Ks}
g_{\text{T},k} \, \delta(\tau-\tau_{\text{T},k})
\, \ab(\theta_{\text{T},k})
\label{eq:ch-T}
\ee
where
$\Ks$ is the number of paths between the terrestrial user 
and the BS, whereas 
$g_{\text{T},k}$, $\tau_{\text{T},k}$, and $\theta_{\text{T},k}$
denote the complex gain, the time delay, and the angle of arrival (AoA) at the BS 
of the $k$th path.
Taking the first receive antenna as reference, 
the vector $\ab(\theta) \in \Cset^J$ in \eqref{eq:ch-T} denotes the response vector 
of the BS in the direction $\theta \in [-\pi/2, \pi/2]$ relative to the
boresight of the ULA and it is given by
\be
\ab(\theta) 
\eqdef \left(1, 
e^{\jmath  \frac{2 \pi}{\lambda_\text{carrier}} d \sin(\theta)}, \ldots, 
e^{\jmath  \frac{2 \pi}{\lambda_\text{carrier}} (J-1) d \sin(\theta)}\right)^{\trasp}  
\ee
where $d$ is
the distance between two adjacent antennas and $\lambda_\text{carrier}=c/f_\text{carrier}$
is the wavelength, with $c$ being the light speed.
From now on, we will set 
$d=\lambda_\text{carrier}/2$ for simplicity.

Since each propagation path is approximately equal to the sum of independent 
micro-scatterers contributions, having same time delay and AoA, the channel gains 
$\{g_{\text{T},k}\}$ 
can thus be modeled as independent zero-mean 
complex Gaussian random variables ({\em Rayleigh fading model}), with variances 
$\sigma_{\text{T},k}^2 \eqdef \Es(|g_{\text{T},k}|^2)$,  for $k \in \{1, 2, \ldots, \Ks\}$.
It can be adopted for $\sigma_{\text{T},k}^2$  the common propagation 
model (see, e.g., \cite{Rap-book})
\be
\sigma_{\text{T},k}^2 = \sigma_{\text{T,ref}}^2 \, 
\pot_\text{T} \, G_\text{T} \, G_\text{BS,T} \, 
 \, \left(\frac{d_{\text{T,ref}}}{d_{\text{T},k}}\right)^{\eta_\text{T}}
\label{eq:sigma_gTk}
\ee
where $\sigma_{\text{T,ref}}^2$ is a unitless constant, 
$\pot_\text{T}$ is the transmit power of the TU, 
$G_\text{T}$ and $G_\text{BS,T}$ represent the antenna gains of 
the TU and the BS, respectively, $d_{\text{T},k}$ denotes the
propagation length along the $k$-th path, 
$d_{\text{T,ref}}$
is a reference distance for the antenna far field, 
and  $\eta_\text{T}$ is the path-loss exponent. 
 
\subsection{Channel model for the aerial user}

For A2C link, we assume a two-ray
channel model,\footnote{The proposed estimation framework can be extended to the multi-ray case  for the A2C channel. However, for blind identification of the 
corresponding channel  parameters, the Doppler shifts of the different rays have to 
fulfill a mild technical condition to avoid ambiguities \cite{Nap-book}.} 
which includes the LoS
path and a scattered (i.e., non-LoS) component \cite{Haas}
(see also Fig.~\ref{fig:fig_1}).
The non-LoS (NLoS) component is physically due to many  
rays reflected or scattered from 
closely-spaced terrestrial obstacles within the
vicinity of the BS,  which are characterized by 
a very narrow beamwidth \cite{Haas}
and appear grouped or ``clustered" in delay
for sufficiently high elevation
angles \cite{Newhall.2003}.
Henceforth, such rays are expected
to have a very similar Doppler shift, as they 
all come from similar directions (narrow beamwidth), and 
a very similar delay (high elevation angle).
Assuming that the UAV is moving (relative to the
BS) at constant {\em radial} speed $v$ within the
observation interval $[0,T_0)$,  with $T_0 \eqdef N_0 \, T_\text{s } \le T_\text{coh}$, 
under the customary assumption that the
communication bandwidth $B \sim 1/T_\text{c}$ is much smaller than the carrier
frequency $f_\text{carrier}$, the far-field low-pass equivalent {\em linear time-varying (LTV)} 
time-delay impulse response of the 
UAV-to-BS channel can be expressed as\footnote{Throughout the
paper, the subscripts L and N
indicate parameters referring to the LoS and NLoS
components of the AU, respectively.} (see, e.g., \cite{Nap-book})
\begin{multline}
\hb_\text{A}(t,\tau)= g_{\text{A,L}} \, e^{\jmath \, 2 \pi  f_{\text{A,L}} t } \, 
\delta(\tau-\tau_{\text{A,L}})
\, \ab(\theta_{\text{A,L}})
\\ + g_{\text{A,N}} \, e^{\jmath \, 2 \pi  f_{\text{A,N}} t } \, \delta(\tau-\tau_{\text{A,N}}) \, \ab(\theta_{\text{A,N}})
\label{eq:ch-A}
\end{multline}
where $g_{\text{A,L}}$ [$g_{\text{A,N}}$],  
$\theta_{\text{A,L}}$ [$\theta_{\text{A,N}}$], 
$\tau_{\text{A,L}}$ [$\tau_{\text{A,N}}$], 
$f_{\text{A,L}} = f_\text{max} \, \cos(\vartheta_{\text{A,L}})$
[$f_{\text{A,N}} = f_\text{max} \, \cos(\vartheta_{\text{A,N}})$]
are the complex (i.e., amplitude and phase) path gain, the AoA at the BS,
the time delay, and  the Doppler shift of the LoS [NLoS] component, with 
$\vartheta_{\text{A,L}} \in [0, \pi]$ [$\vartheta_{\text{A,N}} \in [0, \pi]$] representing the
angle of departure (AoD) of the LoS [NLoS] ray 
relative to the direction of motion
of the UAV and 
$f_\text{max} \eqdef {f_\text{carrier} \, v}/{c} \ll B$.
It is noteworthy that the LTV model \eqref{eq:ch-A} is justified by the fact that
the UAV is allowed to communicate with the BS while it is flying and, hence,
we do not constrain the UAV to transmit only 
in the  hovering-flight state as, e.g., in \cite{Sena}.

Since the speed and physical location of the UAV would change
much slower than the symbol period $T_\text{s}$, the UAV 
related parameters $g_{\text{A,L}}$, $g_{\text{A,N}}$,  
$\theta_{\text{A,L}}$, $\theta_{\text{A,N}}$, 
$\tau_{\text{A,L}}$, $\tau_{\text{A,N}}$,  $f_{\text{A,L}}$,
and $f_{\text{A,N}}$ can be assumed as
unchanged within $N_0 \le N_\text{coh}$ OFDM symbol blocks.
Local stationarity of the scattering geometry is widely used in the
literature and experimentally confirmed in \cite{Va}.
Moreover, we consider an underspread A2C channel, characterized
by $\max\{\tau_{\text{A,L}},\tau_{\text{A,N}}\} \, f_\text{max} \ll 1$, which is 
true for many wireless channels \cite{Proakis}.

The statistical characterization of  \eqref{eq:ch-A} is
herein determined by assuming that the A2C channel is LoS with
probability one.\footnote{It is shown in \cite{Sena} that, 
by using the model and the parameters reported in \cite{ITU,Cherif.2021,Holis.2008}
and assuming a practical UAV altitude range of $25-120$ meters,  
the probability of occurrence of the LoS event between the UAV and the terrestrial BS
is equal to one when either the horizontal distance of the UAV from the BS is 
smaller than about $75$ meters or the elevation angle between the UAV and the BS
is greater than $20$ degrees. 
}
Specifically, the gain  
$g_{\text{A,L}}$ is deterministic and  
takes the value $\sigma_{\text{A,L}} \, e^{j \, \phi_{\text{A,L}}}$. 
The parameter $\phi_{\text{A,L}} \in [0, 2 \pi)$
accounts for a (possible) phase misalignment along the LoS path between
the UAV and the BS.
On the other hand, the NLoS gain $g_{\text{A,N}}$
is modeled as a complex circular zero-mean Gaussian random variable
with variance $\sigma_{\text{A,N}}^2 \eqdef \Es(|g_{\text{A,N}}|^2)$.
Similarly to \eqref{eq:sigma_gTk}, for $\text{C} \in \{\text{L, N}\}$,  
$\sigma_{\text{A,L}}^2$
and $\sigma_{\text{A,N}}^2$ can be modeled as 
\be
\sigma_{\text{A,C}}^2 = \sigma_{\text{A,C,ref}}^2 \, \pot_\text{A} \, G_\text{A} \, G_\text{BS,A,C}
 \, \left(\frac{d_{\text{A,ref}}}{d_{\text{A,C}}}\right)^{\eta_\text{A,C}}
\label{eq:sigma_gA}
\ee
where $\sigma_{\text{A,C,ref}}^2$ is unitless, 
$\pot_\text{A}$ is the transmit power of the AU, 
$G_\text{A}$ and $G_\text{BS,A,C}$ represent the antenna gains of 
the AU and the BS, respectively, $d_{\text{A,C,ref}}$
is a reference distance for the antenna far field, 
$d_{\text{A,C}}$ denotes the propagation length of path 
$\text{C} \in \{\text{L, N}\}$, and 
$\eta_\text{A,C}$ is the corresponding path-loss exponent. 
It should be observed that the UAV-to-BS channel 
is described by a {\em Rician fading model}, whose
corresponding Rician factor is 
$K_\text{A} \eqdef \sigma_{\text{A,L}}^2/\sigma_{\text{A,N}}^2$.
We underline that
the path $\text{C} \in \{\text{L, N}\}$ of the 
AU and the MPCs of the TU see different BS antenna gains 
$G_\text{BS,A,C}$ and  $G_\text{BS,T}$, respectively,
which reflects the fact that the AU and TU  
may be served by different lobes of the BS antennas.

\subsection{Receive signal model before CP removal}

Hereinafter, without loss of generality,
the reported signal models refer to the observation interval 
 $[0,T_0)$ of the TU channel, which corresponds to 
the transmission of $N_0$ OFDM symbols. 
After filtering, for $t \in [0,T_0)$ and $j \in \{1,2,\ldots, J\}$,
the baseband received signal
at the $j$-th antenna of the BS can be written as 
\begin{multline}
\y_j(t) = g_{\text{A,L}} \, e^{\jmath \, 2 \pi  f_{\text{A,L}} t } \, \xa(t-\tau_{\text{A,L}}) \, 
e^{\jmath \pi (j-1) \sin(\theta_{\text{A,L}})}
\\ + 
g_{\text{A,N}} \, e^{\jmath \, 2 \pi  f_{\text{A,N}} t } \, \xa(t-\tau_{\text{A,N}}) \, 
e^{\jmath \pi (j-1) \sin(\theta_{\text{A,N}})}
\\ +
\sum_{k=1}^{\Ks} g_{\text{T},k}  \, \xs(t-\tau_{\text{T},k}) \, 
e^{\jmath \pi (j-1) \sin(\theta_{\text{T},k})} + w_j(t)
\label{eq:sig}
\end{multline}
where $w_j(t)$ is the complex envelope of (filtered) noise statistically 
independent of $x_\text{TX}(t)$, for $\text{TX} \in \{\text{A}, \text{T}\}$.

If we denote with $\psi_\text{ADC}(t)$
the impulse response of the (anti-aliasing) filter at the
input of the analog-to-digital converter (ADC) at the BS, the
impulse response  of the cascade of the
DAC interpolation filter and the ADC antialiasing
filter is given by $\psi(t) \eqdef \psi_\text{DAC}(t) \ast \psi_\text{ADC}(t)$,
obeying
$\psi(t) \equiv 0$ for $t \not \in [0, L_\text{filter}\, T_\text{c})$.
The pulse $\psi(t)$
and its finite duration $L_\text{filter} \, T_\text{c}$ are known at the BS.
A very common assumption in multicarrier applications is
that
$L_\text{filter} \, T_\text{c} +\tau_\text{max} < T_\text{s}$,
with 
$\tau_\text{max} \eqdef \max(\tau_{\text{A,L}}, \tau_{\text{A,N}}, 
\tau_{\text{T},1}, \ldots, \tau_{\text{T},\Ks})$.
Moreover, we customarily assume that the BS
has been previously locked to 
the multipath component 
at (approximately) the minimum delay
$\tau_\text{min} \eqdef \min(\tau_{\text{A,L}},  \tau_{\text{A,N}}, 
\tau_{\text{T},1}, \ldots, \tau_{\text{T},\Ks})$
and, without
loss of generality, we set
$\tau_\text{min}=0$. 

The signal \eqref{eq:sig} is sampled at 
$t_{n,p} \eqdef n T_\text{s} + p \, T_\text{c}$, for $n \in \Zset$ and $p \in \{0,1\ldots, P-1\}$.
Let  $y_j^{[p]}[n] \eqdef y_j(t_{n,p})$ be the discrete-time
counterpart of \eqref{eq:sig}, for $n \in \{0,1,\ldots,N_0-1\}$, one gets
\begin{multline}
y_j^{[p]}[n] =
g_{\text{A,L}} \, e^{\jmath \, 2 \pi \nu_{\text{A,L}} \left(n +\frac{p}{P} \right) } \,
x_{\text{A,L}}^{[p]}[n] \, e^{\jmath \pi (j-1) \sin(\theta_{\text{A,L},})} \\ +
g_{\text{A,N}} \, e^{\jmath \, 2 \pi \nu_{\text{A,N}} \left(n +\frac{p}{P} \right) } \,
x_{\text{A,N}}^{[p]}[n] \, e^{\jmath \pi (j-1) \sin(\theta_{\text{A,N},})}  \\ +
\sum_{k=1}^{\Ks} g_{\text{T},k} \,
\xsk^{[p]}[n] \, e^{\jmath \pi (j-1) \sin(\theta_{\text{T},k})} + w_j^{[p]}[n]
\label{eq:sig-ric}
\end{multline}
where
$\nu_{\text{A,C}} \eqdef f_{\text{A,C}} \, T_\text{s} \in [-1/2, 1/2)$ 
is the {\em normali\-zed} Doppler shift of the UAV relative to 
path  $\text{C} \in \{\text{L, N}\}$,  whereas
\barr
x_\text{A,C}^{[p]}[n] & = 
\sum_{\ell=n-1}^{n} \sum_{q=0}^{P-1}
u_{\text{A}}^{[q]}[\ell] 
\nonumber 
\\ & \hspace{15mm} \cdot 
\alpha_{\text{A,C}} 
\left[(n-\ell) P +(p-q) -d_{\text{A,C}}\right] 
\\
x_{\text{T},k}^{[p]}[n] & = 
\sum_{\ell=n-1}^{n} \sum_{q=0}^{P-1}
u_{\text{T}}^{[q]}[\ell]  
\nonumber 
\\ & \hspace{15mm} \cdot 
\alpha_{\text{T},k} 
\left[(n-\ell) P +(p-q) -d_{\text{T},k}\right] 
\label{eq:xpn}
\earr
where
\barr
\alpha_{\text{A,C}}[h] & \eqdef \psi(h \, T_\text{c}-\chi_{\text{A,C}})
\quad (h \in \mathbb{Z})
\\
\tau_{\text{A,C}} & = d_{\text{A,C}} \, T_\text{c} + \chi_{\text{A,C}}
\\
\alpha_{\text{T},k}[h] & \eqdef \psi(h \, T_\text{c}-\chi_{\text{T},k})
\\
\tau_{\text{T},k} & = d_{\text{T},k} \, T_\text{c} + \chi_{\text{T},k}
\earr
with integer delays $d_{\text{A,C}}$ and $d_{\text{T},k}$,  
and fractional delays $\chi_{\text{A,C}} \in [0, T_\text{c})$
and  $\chi_{\text{T},k} \in [0, T_\text{c})$,
and $w_j^{[p]}[n] \eqdef w_j(t_{n,p})$.

By gathering the obtained samples of the received signal
into 
$\overline{\yb}_j[n] \eqdef (y_j^{[0]}[n], y_j^{[1]}[n],\ldots,y_j^{[P-1]}[n])^\trasp \in \Cset^P$ and accounting
for \eqref{eq:sig-ric}, we obtain the following vector model 
\begin{multline}
\overline{\yb}_j[n] = \Hatildezero[n] \, \bab[n] + \Hatildeuno[n] \, \bab[n-1]
\\ +
\Hstildezero \, \bsb[n]  + \Hstildeuno \, \bsb[n-1] + \overline{\wb}_j[n]
\label{eq:y-IBI}
\end{multline}
for $n \in \{0,1,\ldots,N_0-1\}$ and $j \in \{1,2,\ldots, J\}$, where
\barr
\overline{\mathbf{H}}_{\text{A},j}^{(b)} [n] & \eqdef 
g_{\text{A,L}}  \left( \mathbf{D}_{\text{A,L}} \,
\mathbf{T}_{\text{A,L}}^{(b)} \, \bm{\Omega}_\text{A} \right) e^{\jmath \, 2 \pi \nu_{\text{A,L}} n}
\, e^{\jmath \pi (j-1) \sin(\theta_{\text{A,L}})} 
\nonumber \\ & +
g_{\text{A,N}}  \left( \mathbf{D}_{\text{A,N}} \,
\mathbf{T}_{\text{A,N}}^{(b)} \, \bm{\Omega}_\text{A} \right) 
e^{\jmath \, 2 \pi \nu_{\text{A,N}} n}
\, e^{\jmath \pi (j-1) \sin(\theta_{\text{A,N}})} 
\label{eq:mat-A}
\\
\overline{\mathbf{H}}_{\text{T},j}^{(b)} & \eqdef  \sum_{k=1}^{K_\text{T}}
g_{\text{T},k} \left(\mathbf{T}_{\text{T},k}^{(b)} \, \bm{\Omega}_\text{T} \right) e^{\jmath \pi (j-1) \sin(\theta_{\text{T},k})}
\label{eq:mat-T}
\earr
for $b \in \{0,1\}$, with
\be
\mathbf{D}_{\text{A,C}} \eqdef \diag(1, e^{\jmath \, \frac{2 \pi}{P} \nu_{\text{A,C}}},
\ldots, e^{\jmath \, \frac{2 \pi}{P} \nu_{\text{AC}} (P-1)}) 
\ee
for $\text{C} \in \{\text{L, N}\}$, 
whereas the $(p+1,q+1)$th entry of the Toepltiz matrices
$\mathbf{T}_{\text{A,C}}^{(b)}\in \Rset^{P \times P}$
and $\mathbf{T}_{\text{T},k}^{(b)}\in \Rset^{P \times P}$ is given by
$\alpha_{\text{A,C}}\left[b\,  P +(p-q) -d_{\text{A,C}}\right]$
and $\alpha_{\text{T},k}\left[b\,  P +(p-q) -d_{\text{T},k}\right]$,
respectively,  for $p, q \in \{0,1\ldots, P-1\}$,
\be
\bm{\Omega}_\text{TX} \eqdef  \bm{I}_\text{cp} \, \bm{W}_M \in \Cset^{P \times M}
\ee
and
$\overline{\wb}_j[n] \eqdef (w^{[0]}_j[n], w^{[1]}_j[n],\ldots,w^{[P-1]}_j[n])^\trasp \in \Cset^P$
is a zero-mean complex circular Gaussian random vector
with $\Es(\overline{\wb}_{j_1}[n_1] \, \overline{\wb}_{j_2}^\herm[n_2])=
\sigma_w^2 \, \delta_{n_1-n_2} \, \delta_{j_1-j_2} \, \I_{P}$.

The signal model \eqref{eq:y-IBI} {\em before} CP removal is used 
in Subsection~III-A to blindly
estimate the Doppler shifts and the time delays of the AU.

\subsection{Receive signal model after CP removal}
For $n \in \{0,1,\ldots,N_0-1\}$ and $j \in \{1,2,\ldots, J\}$,
interblock interference (IBI) in \eqref{eq:sig-ric} can be perfectly suppressed
through CP removal, thus yielding, after DFT, 
\barr
\yb_j[n] & \eqdef \bm{W}_M^\herm \, \bm{R}_\text{cp} \, \overline{\yb}_j[n] 
\nonumber \\ & = \bm{H}_{\text{A},j}[n] \, \bab[n] +
\bm{H}_{\text{T},j} \, \bsb[n]  + \wb_j[n]
\label{eq:y_j}
\earr
provided that 
$\Lcp \ge L_\text{filter}+\lfloor\tau_\text{max}/T_\text{c}\rfloor$,
where $\bm{R}_\text{cp} \in \mathbb{R}^{M \times P}$ performs CP
removal,
$\bm{H}_{\text{A},j}[n] \eqdef \bm{W}_M^\herm \, \bm{R}_\text{cp} \,
\overline{\mathbf{H}}_{\text{A},j}^{(0)} [n] \in \Cset^{M \times M}$
and $\bm{H}_{\text{T},j}\eqdef \bm{W}_M^\herm \, \bm{R}_\text{cp} \,
\overline{\mathbf{H}}_{\text{T},j}^{(0)} \in \Cset^{M \times M}$,
and, finally, 
the noise contribution is $\wb_j[n] \eqdef \bm{W}_M^\herm \, \bm{R}_\text{cp} \, \overline{\wb}_j[n] \in \Cset^M$.
Moreover, accounting for \eqref{eq:mat-T},
the matrix $\bm{H}_{\text{T},j}$ can be written as
reported at the top of the next page in \eqref{eq:HJexpl},
\begin{figure*}[!t]
\normalsize
\be
\bm{H}_{\text{T},j}   =  \bm{W}_M^\herm \sum_{k=1}^{K_\text{T}}
\underbrace{\left[ g_{\text{T},k} \, e^{\jmath \pi (j-1) \sin(\theta_{\text{T},k})} 
\sum_{\ell=0}^{\Lf} \alpha_{\text{T},k}[\ell]  \, 
\bm{R}_\text{cp} \, \Fb^{\ell+d_{\text{T},k}} \, \bm{I}_\text{cp} \right]}_{\bm{W}_M \, 
\Mb_{\text{T},j,k}\, \bm{W}_M^\herm} \bm{W}_M = \Mb_{\text{T},j}
\label{eq:HJexpl}
\ee
\hrulefill
\end{figure*}
where the forward matrix $\Fb \in \Cset^{P \times P}$ has been defined in 
Subsection~I-D
and we have observed that the matrix in square bra\-ckets is circulant and, thus, it can be 
equivalently written as $\bm{W}_M \, 
\Mb_{\text{T},j,k} \bm{W}_M^\herm$, with
\be
\Mb_{\text{T},j} \eqdef \sum_{k=1}^{K_\text{T}} \Mb_{\text{T},j,k} \:.
\ee
The matrix 
$\Mb_{\text{T},j,k} \eqdef \diag(\boldsymbol{\mu}_{\text{T},j,k})$ is diagonal, with
\be
\boldsymbol{\mu}_{\text{T},j,k} \eqdef \sqrt{M} \, \bm{W}_M^\herm \, \Lb \, \gb_{\text{T},j,k}
\in \Cset^M
\label{eq:mu-TU}
\ee
where
\barr
\Lb & \eqdef (\I_{\Lcp}, \bm{O}_{(M-\Lcp) \times \Lcp}^\trasp)^\trasp \in \Rset^{M \times \Lcp}
\\
\gb_{\text{T},j,k} & \eqdef (\bm{0}_{d_{\text{T},k}}^\trasp, 
g_{\text{T},k} \, e^{\jmath \pi (j-1) \sin(\theta_{\text{T},k})} \, \boldsymbol{\alpha}_{\text{T},k}^\trasp,
\\ & \hspace{25mm} \ldots
\bm{0}_{\Lcp-\Lf-d_{\text{T},k}-1}^\trasp)^\trasp \in \Cset^{\Lcp}
\\
\boldsymbol{\alpha}_{\text{T},k} & \eqdef (\alpha_{\text{T},k}[0], \alpha_{\text{T},k}[1], \ldots,
\alpha_{\text{T},k}[\Lf])^\trasp \in \Rset^{\Lf+1} \: .
\label{eq:alphabold}
\earr
On the other hand, the transmission of the AU is 
adversely affected by ICI 
due to the presence of Doppler shifts: in general, the channel matrix $\mathbf{H}_{\text{A},j} [n]$ 
cannot be diagonalized through DFT.

By collecting the data received by all the antennas of the BS
in the vector $\yb[n] \eqdef (\yb_1^\trasp[n], \yb^\trasp_2[n], \ldots, \yb^\trasp_J[n])^\trasp 
\in \Cset^{J M}$, we obtain from \eqref{eq:y_j} and \eqref{eq:HJexpl} the signal model
\be
\yb[n] \eqdef \bm{H}_{\text{A}}[n] \, \bab[n] +
\bm{M}_{\text{T}} \, \bsb[n]  + \wb[n]
\label{eq:y}
\ee
for $n \in \{0,1,\ldots,N_0-1\}$, where 
\barr
\bm{H}_{\text{A}}[n] & \eqdef (\bm{H}_{\text{A},1}^\trasp[n], 
\bm{H}_{\text{A},2}^\trasp[n],
\ldots, \bm{H}_{\text{A},J}^\trasp[n])^\trasp \in \Cset^{(J M) \times M}
\\
\bm{M}_{\text{T}} & \eqdef (\bm{M}_{\text{T},1}^\trasp, 
\bm{M}_{\text{T},2}^\trasp,
\ldots, \bm{M}_{\text{T},J}^\trasp)^\trasp \in \Cset^{(J M) \times M}
\earr
and $\wb[n] \eqdef (\wb_1^\trasp[n], \wb^\trasp_2[n], \ldots, \wb^\trasp_J[n])^\trasp \in \Cset^{J M}$. 


\subsection{Nonorthogonal pilot allocation scheme}

To acquire the channel parameters needed for demodulating the 
data (i.e., information-bearing) symbols transmitted by the users, 
the BS might rely on training (or pilot) symbols sent by both the AU and TU, 
over reserved time-frequency RBs in uplink.
Specifically, we assume that user $\text{TX} \in \{\text{A}, \text{T}\}$
inserts $1 \le Q_\text{TX} \le M$ pilot symbols at
distinct subcarrier locations 
\begin{multline}
\mathcal{M}_\text{TX} \eqdef
\{m_{\text{TX},0}, m_{\text{TX},1}, 
\ldots, m_{\text{TX},Q_\text{TX}-1}\} 
\\
\subseteq \{0,1,\ldots, M-1\}
\end{multline}
during 
the $1 \le N_\text{TX,train} \ll N_0$ distinct
symbol intervals
\begin{multline}
 \mathcal{N}_\text{TX} \eqdef \{n_{\text{TX},1}, n_{\text{TX},2}, 
\ldots, n_{\text{TX}, N_\text{TX, train}}\} 
\\
\subset \{0,1,\ldots, N_0-1\} \: .
\end{multline}

In NOMA schemes, 
the training phase of one user suffers from 
the interference generated by either the pilots 
or the data transmitted by the other one. 
Perfect separation of pilot symbols at the BS
is possible only if the TU will not transmit 
neither pilots nor data on the set 
$\mathcal{M}_\text{A}$ of subcarriers
over the OFDM signaling intervals  $\mathcal{N}_\text{A}$ and vice versa. 
Obviously, besides requiring a stringent protocol-level cooperation between 
the aerial and terrestrial networks, such a latter strategy comes 
at a (possibly unaffordable) cost of spectral efficiency
due to the time-varying nature of the AU channel 
and it is not considered herein.

\section{Channel estimation for the aerial user}
\label{sec:estimation-AU}

The task of the AU channel estimator is to provide to the WL-TV SIC filter
the channel matrices $\bm{H}_{\text{A},1}[n], \bm{H}_{\text{A},2}[n],\ldots,
\bm{H}_{\text{A},J}[n]$, for $n \in \{0,1,\ldots,N_0-1\}$.
In this case, since the time-varying nature of the A2C channel prevents 
the diagonalization of $\bm{H}_{\text{A},j}[n]$ through DFT, 
it is not possible to separate pilots and data symbols in the 
frequency-domain. For such a reason, we set  
$\mathcal{M}_\text{A} \equiv \{0,1,\ldots, M-1\}$ or,
equivalently, $Q_\text{A}=M$, i.e., 
the AU inserts pilots at all the subcarriers within  
the symbol intervals  specified by 
$\mathcal{N}_\text{A}=\{n_{\text{A},1}, n_{\text{A},2},  \ldots, n_{\text{A},N_\text{A,train}}\}$.

To reduce the amount of training, we 
propose a semi-blind approach to estimate 
the unknowns characterizing
the channel matrices $\bm{H}_{\text{A},1}[n], \bm{H}_{\text{A},2}[n],\ldots,
\bm{H}_{\text{A},J}[n]$. Speci\-fically, Doppler and delay 
parameters are blindly estimated  by exploiting the unique
ACS properties of the signal transmitted by the AU,
without relying on pilots.
On the other hand, complex gain and 
AoA parameters 
are estimated by using the 
pilot symbols known by the BS.
The estimators derived herein are based on the
assumption that the AU channel parameters 
are deterministic but unknown quantities.

\begin{figure*}[t]
\centering
\includegraphics[width=1.8\columnwidth]{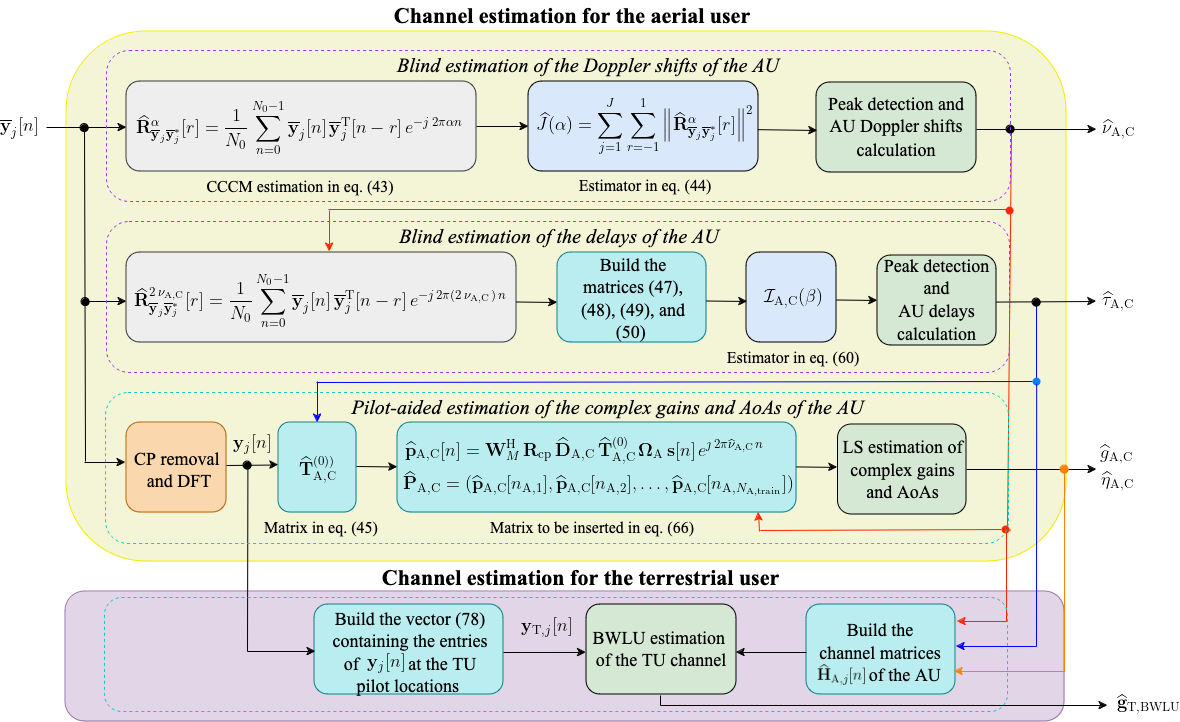}
\caption{A summary of the proposed channel estimators for both the AU and TU
(the BWLU estimator for the TU channel is depicted only).}
\label{fig:fig_sum_alg}
\end{figure*}

The proposed estimators of the AU channel parameters 
are summarized in Fig.~\ref{fig:fig_sum_alg}.

\subsection{Blind estimation of Doppler and delay parameters}
\label{sec:Doppler-gain}

The starting point of our blind estimation procedure is the signal 
\eqref{eq:y-IBI}, i.e.,
the \textit{entire} OFDM block 
(including the CP) is processed
for acquiring Doppler and delay
parameters of the AU transmission.
The ACS features of \eqref{eq:y-IBI}
stem from
the fact that $\overline{\mathbf{H}}_{\text{A},j}^{(b)} [n]$
in \eqref{eq:mat-A} is a discrete-time
{\em almost-periodic} (AP) matrix \cite{Cord} 
with (possibly)
incommensurate frequencies 
$\{\nu_{\text{A,L}}, {\nu}_{\text{A,N}}\} \subseteq [-1/2, 1/2)$.
Consequently, for $r \in \{-1,0,1\}$ and $j \in \{1,2,\ldots, J\}$,
the correlation matrix
$\Rb_{\overline{\yb}_j\overline{\yb}_j}[n, r] \eqdef \Es(\overline{\yb}_j[n] \,
\overline{\yb}_j^\herm[n-r]) \in \Cset^{P \times P}$
and its conjugate counterpart $\Rb_{\overline{\yb}_j \overline{\yb}_j^*}[n, r] \eqdef 
\Es(\overline{\yb}_j[n] \, \overline{\yb}_j^\trasp[n-r]) \in \Cset^{P \times P}$
are AP matrices, too, 
and the multivariate process \eqref{eq:y-IBI} is said to be
second-order wide-sense ACS \cite{Gardner-book}.

Capitalizing on the fact that the AU transmits noncircular symbols
while the TU adopts a circular modulation format (modulation diversity), 
the proposed blind estimation approach relies only on the conjugate second-order statistics
of the ACS random process \eqref{eq:y-IBI}, which
turn out to be free from both TU and noise contributions (a consequence of Doppler diversity). 
For $j \in \{1,2,\ldots, J\}$, one gets
\begin{multline}
\Rb_{\overline{\yb}_j\overline{\yb}_j^{*}}[n, r] =
\sum_{\text{C}_1 \in \{\text{L, N}\}} \sum_{\text{C}_2 \in \{\text{L, N}\}}
\bm{\Xi}_{\text{A},j,\text{C}_1,\text{C}_2}[r] 
\\ \cdot e^{\jmath 2 \pi (\nu_{\text{A},\text{C}_1} 
+\nu_{\text{A},\text{C}_2}) n}
\label{eq:Ryy}
\end{multline}
where  the $P \times P$ matrices in \eqref{eq:Ryy} are defined
in \eqref{eq:Psi-1}--\eqref{eq:Psi+1} at the top of the next page,
with $\bm{\Delta} \in \Cset^{M \times M}$ being 
a {\em known} diagonal matrix derived in the Appendix
for the $\pi/2$-BPSK case
and depending on the noncircularity 
property of the AU symbols.
\begin{figure*}[!t]
\normalsize
\barr
\bm{\Xi}_{\text{A},j,\text{C}_1,\text{C}_2}[-1]  & \eqdef  g_{\text{A},\text{C}_1} \, 
e^{\jmath \pi (j-1) \sin(\theta_{\text{A},\text{C}_1})} \, 
g_{\text{A},\text{C}_2} \, 
e^{\jmath \pi (j-1) \sin(\theta_{\text{A},\text{C}_2})}
\, e^{\jmath \, 2 \pi \nu_{\text{A},\text{C}_2}}  \,
(\mathbf{D}_{\text{A},\text{C}_1} \, \mathbf{T}_{\text{A},\text{C}_1}^{(0)} \, 
\bm{\Omega}_\text{A}) \, \bm{\Delta}
\, (\mathbf{D}_{\text{A},\text{C}_2} \, \mathbf{T}_{\text{A},\text{C}_2}^{(1)} \, 
\bm{\Omega}_\text{A})^\trasp
\label{eq:Psi-1}
\\
\bm{\Xi}_{\text{A},j,\text{C}_1,\text{C}_2}[0] & \eqdef 
g_{\text{A},\text{C}_1} \, 
e^{\jmath \pi (j-1) \sin(\theta_{\text{A},\text{C}_1})} \, 
g_{\text{A},\text{C}_2} \, 
e^{\jmath \pi (j-1) \sin(\theta_{\text{A},\text{C}_2})}
\, 
 \mathbf{D}_{\text{A},\text{C}_1} \, [(\mathbf{T}_{\text{A},\text{C}_1}^{(0)} 
 \, \bm{\Omega}_\text{A}) \, \bm{\Delta} \, 
(\mathbf{T}_{\text{A},\text{C}_2}^{(0)} \, \bm{\Omega}_\text{A})^\trasp
\nonumber \\ & \hspace{80mm}
+ (\mathbf{T}_{\text{A},\text{C}_1}^{(1)} \, \bm{\Omega}_\text{A}) \, \bm{\Delta} \, 
(\mathbf{T}_{\text{A},\text{C}_2}^{(1)} \, \bm{\Omega}_\text{A})^\trasp ] \, 
\mathbf{D}_{\text{A},\text{C}_2} 
\label{eq:Psi-0}
\\
\bm{\Xi}_{\text{A},j,\text{C}_1,\text{C}_2}[1] & \eqdef
g_{\text{A},\text{C}_1} \, 
e^{\jmath \pi (j-1) \sin(\theta_{\text{A},\text{C}_1})} \, 
g_{\text{A},\text{C}_2} \, 
e^{\jmath \pi (j-1) \sin(\theta_{\text{A},\text{C}_2})}
 \, e^{-\jmath \, 2 \pi \nu_{\text{A},\text{C}_2}} \,
(\mathbf{D}_{\text{A},\text{C}_1} \, \mathbf{T}_{\text{A},\text{C}_1}^{(1)} \, \bm{\Omega}_\text{A}) \, \bm{\Delta} \, 
(\mathbf{D}_{\text{A},\text{C}_2} \, \mathbf{T}_{\text{A},\text{C}_2}^{(0)} \, \bm{\Omega}_\text{A})^\trasp 
\label{eq:Psi+1}
\earr
\hrulefill
\end{figure*}
Under mild regularity conditions, the AP matrix \eqref{eq:Ryy} can be equivalently written
in terms of the generalized Fourier series expansion 
as follows
\be
\Rb_{\overline{\yb}_j\overline{\yb}^{*}_j}[n,r] =
\sum_{\alpha}
\Rb_{\overline{\yb}_j\overline{\yb}_j^{*}}^{\alpha}[r] \, e^{\jmath 2 \pi \alpha n}
\ee
where, for $r \in \{-1,0,1\}$,
\barr
\Rb_{\overline{\yb}_j\overline{\yb}_j^{*}}^{\alpha}[r] & \eqdef
\left \langle \Rb_{\overline{\yb}_j\overline{\yb}_j^{*}}[n,r]
\, e^{-j 2 \pi \alpha n} \right \rangle
\nonumber \\ & =
\lim_{N \to +\infty}
\frac{1}{2 N+1} \sum_{n=-N}^{N} \Rb_{\overline{\yb}_j\overline{\yb}_j^{*}}[n,r]
\, e^{-j 2 \pi \alpha n}
\label{eq:Ralphagen}
\earr
is the 
{\em conjugate cyclic correlation matrix (CCCM)}
at cycle frequency $\alpha$ and it is given by
\be
\Rb_{\overline{\yb}_j\overline{\yb}_j^{*}}^{\alpha}[r]  =
\begin{cases}
\bm{\Xi}_{\text{A},j,\text{L},\text{L}}[r] \:, & \text{$\alpha=2 \, \nu_{\text{A,L}}$} \:;
\\
\bm{\Xi}_{\text{A},j,\text{N},\text{N}}[r] \:, & \text{$\alpha=2 \, \nu_{\text{A,N}}$} \:;
\\
\bm{\Xi}_{\text{A},j,\text{L},\text{N}}[r]+\bm{\Xi}_{\text{A},j,\text{N},\text{L}}[r] \:, 
& \text{$\alpha=\nu_{\text{A,L}}+\nu_{\text{A,N}}$} \:;
\\
\Zero_{P \times P} \:, & \text{otherwise} \: .
\end{cases}
\label{eq:Ralpha}
\ee
It is worth noticing that there are three distinct cycle
frequencies 
$2 \, \nu_{\text{A,L}}$, 
$2 \, \nu_{\text{A,N}}$, and
$\nu_{\text{A,L}}+\nu_{\text{A,N}}$,\footnote{Since the LoS and NLoS components
arrive at the BS from different angular directions, one can safely 
assume that $\nu_{\text{A,L}} \neq \nu_{\text{A,N}}$.} which arise 
from linear combinations of the Doppler shifts 
$\nu_{\text{A,L}}$ and $\nu_{\text{A,N}}$. Thus,  
the cycle frequency set  of \eqref{eq:Ryy} is 
\be
\mathcal{A} =
\left\{2 \, \nu_{\text{A,L}}, 2\, \nu_{\text{A,N}}, \nu_{\text{A,L}}+\nu_{\text{A,N}}\right\}\:.
\label{eq:Ayy}
\ee

Another  key observation underlying our blind estimation approach
is that, in practice, $\Rb_{\overline{\yb}_j\overline{\yb}_j^*}^{\alpha}[r]$
can be directly estimated from the received data 
using the consistent estimate (see, e.g., \cite{Dand})
\be
\widehat{\Rb}_{\overline{\yb}_j\overline{\yb}_j^*}^{\alpha}[r] =
\frac{1}{N_0} \sum_{n=0}^{N_0-1}
\overline{\yb}_j[n] \,  \overline{\yb}_j^\trasp[n-r] \, e^{-j \, 2 \pi \alpha n}\: .
\label{eq:Rciclest}
\ee

\subsubsection{Blind estimation of the Doppler shifts}
\label{sec:Doppler}

Estimation
of the Doppler shifts $f_{\text{A,L}}$ and $f_{\text{A,N}}$
is equivalent to that of $\nu_{\text{A,L}}$ and 
$\nu_{\text{A,N}}$, respectively.
For $j \in \{1,2,\ldots, J\}$,
knowledge of the CCCMs $\Rb_{\overline{\yb}\overline{\yb}^*}^{\alpha}[-1]$,
$\Rb_{\overline{\yb}_j\overline{\yb}_j^*}^{\alpha}[0]$, 
and $\Rb_{\overline{\yb}_j\overline{\yb}_j^*}^{\alpha}[1]$ enables one 
to blindly retrieve
the unknown cycle frequency set \eqref{eq:Ayy}
through the {\em one-dimensional} function:
\begin{multline}
J(\alpha) \eqdef \sum_{j=1}^J
\sum_{r=-1}^{1} \left \|\Rb_{\overline{\yb}_j\overline{\yb}_j^*}^{\alpha}[r] \right \|^2
\\ =
\sum_{j=1}^J \sum_{r=-1}^{1} \trace\left(\Rb_{\overline{\yb}_j\overline{\yb}_j^*}^{\alpha}[r]
\left\{\Rb_{\overline{\yb}_j\overline{\yb}_j^*}^{\alpha}[r]\right\}^\herm \right)\: ,
\\ \quad \text{with $\alpha \in [-1/2, 1/2)$} \: .
\label{eq:J}
\end{multline}
By using differential calculus arguments,
it can be shown
that the function
$J(\alpha)$ has three local maxima 
in $[-1/2,1/2)$ at points
$2 \, \nu_{\text{A,L}}$, $2\, \nu_{\text{A,N}}$,  and 
$\nu_{\text{A,L}}+\nu_{\text{A,N}}$.
Hence, the cycle frequencies
of the second-order ACS
process \eqref{eq:y-IBI}  can be acquired by  searching
for the maxima of \eqref{eq:J}
over $[-1/2,1/2)$, provided that 
$|\nu_{\text{A,L}}|, |\nu_{\text{A,N}}| \le 1/4$.
Specifically, if $\nu_{\text{A,L}} < \nu_{\text{A,N}}$, 
the maximum points of $J(\alpha)$ in $[-1/2,1/2)$
are $2 \, \nu_{\text{A,L}}< \nu_{\text{A,L}}+\nu_{\text{A,N}}
< 2\, \nu_{\text{A,N}}$ in increasing order. On the contrary,  
if $\nu_{\text{A,L}} > \nu_{\text{A,N}}$, 
the increasing order of the maximum points 
is  $2 \, \nu_{\text{A,L}}< \nu_{\text{A,L}}+\nu_{\text{A,N}}
< 2\, \nu_{\text{A,L}}$.
Therefore, by picking up the smallest and largest maximum points of 
$J(\alpha)$ in $[-1/2,1/2)$, one can acquire the Doppler shifts of the AU
up to a permutation ambiguity, i.e., it is not possible to infer if the smallest (or, conversely, largest)
maximum point of \eqref{eq:J} is equal to $2 \, \nu_{\text{A,L}}$
or $2 \, \nu_{\text{A,N}}$. However, such an ambiguity is irrelevant
for the detection process of the symbols transmitted by the AU.

In practice, the estimates of the Doppler shifts 
can be obtained from data by finding the peaks of the function
$\widehat{J}(\alpha)$ obtained from \eqref{eq:J} by replacing 
\eqref{eq:Ralphagen} with \eqref{eq:Rciclest}.

\subsubsection{Blind estimation of the time delays}
\label{sec:delay}
Hereinafter, we assume that the Doppler shifts have been previously acquired as proposed in 
Subsection~III-A-1 and, for the time being, they are
assumed perfectly known.

The proposed estimation technique to acquire   
the delays
$\tau_{\text{A,L}}= d_{\text{A,L}} \, T_\text{c} + \chi_{\text{A,L}}$
and $\tau_{\text{A,N}}= d_{\text{A,N}} \, T_\text{c} + \chi_{\text{A,N}}$
relies again on  the AP
conjugate correlation matrices
$\Rb_{\overline{\yb}_j\overline{\yb}_j^*}[n,-1]$,
$\Rb_{\overline{\yb}_j\overline{\yb}_j^*}[n,0]$, and 
$\Rb_{\overline{\yb}_j\overline{\yb}_j^*}[n,1]$
in \eqref{eq:Ryy}. In addition, we capita\-lize on the following
parameterization of the Toeplitz matrices
$\mathbf{T}_{\text{A,L}}^{(b)}$ and $\mathbf{T}_{\text{A,N}}^{(b)}$ - defined in
 \eqref{eq:mat-A} with $b \in \{0,1\}$ - in terms of 
the  forward shift $\Fb \in \Rset^{P \times P}$
and backward shift $\Bb \in \Rset^{P \times P}$ matrices 
(see Subsection~I-D):
\barr
\mathbf{T}_{\text{A,C}}^{(0)} & = \sum_{\ell=0}^{\Lcp}
\psi(\ell \, T_\text{c}-\tau_{\text{A,C}}) \, \Fb^{\ell}
\label{eq:T0}
\\
\mathbf{T}_{\text{A,C}}^{(1)} & =
\sum_{\ell=1}^{\Lcp}
\psi(\ell \, T_\text{c}-\tau_{\text{A,C}}) \, \Bb^{P-\ell}
\label{eq:T1}
\earr
for $\text{C} \in \{\text{L, N}\}$, 
where the pulse $\psi(t)$ is known
at the BS.

Starting from the CCCMs \eqref{eq:Ralpha}, for $r \in\{-1,0,1\}$,
at cycle frequency $\alpha=2 \, \nu_{\text{A,C}}$, which can 
be obtained from \eqref{eq:Psi-1}--\eqref{eq:Psi+1} by setting 
$\text{C}_1=\text{C}_2=\text{C} \in \{\text{L, N}\}$, we build the matrices
\barr
\bm{\Phi}_{\text{A},j,\text{C}}[-1] & \eqdef  e^{-j 2 \pi \nu_{\text{A,C}}}  \,
\mathbf{D}_{\text{A,C}}^* \, \Rb_{\overline{\yb}_j\overline{\yb}_j^*}^{2 \,
\nu_{\text{A,C}}}[-1] \, \mathbf{D}_{\text{A,C}}^*
\nonumber \\ & =  
g_{\text{A,C}}^2 \, e^{\jmath 2 \pi (j-1) \sin(\theta_{\text{A,C}})} 
\nonumber \\ & \hspace{20mm} \cdot 
(\mathbf{T}_{\text{A,C}}^{(0)} \,
\bm{\Omega}_\text{A}) \, \bm{\Delta} \, (\mathbf{T}_{\text{A,C}}^{(1)} \,
\bm{\Omega}_\text{A})^\trasp 
\label{eq:Phi-1}
\\
\bm{\Phi}_{\text{A},j,\text{C}}[0] & \eqdef
\mathbf{D}_{\text{A,C}}^* \, \Rb_{\overline{\yb}_j\overline{\yb}_j^*}^{2 \, \nu_{\text{A,C}}}[0]
\, \mathbf{D}_{\text{A,C}}^*
\nonumber \\ & = 
 g_{\text{A,C}}^2 \, e^{\jmath 2 \pi (j-1) \sin(\theta_{\text{A,C}})}
\nonumber \\ & \cdot 
\left[(\mathbf{T}_{\text{A,C}}^{(0)} \, \bm{\Omega}_\text{A}) \, \bm{\Delta} \, 
(\mathbf{T}_{\text{A,C}}^{(0)} \, \bm{\Omega}_\text{A})^\trasp
\right. \nonumber \\ & \left. \hspace{15mm}
+ (\mathbf{T}_{\text{A,C}}^{(1)} \, \bm{\Omega}_\text{A}) \, \bm{\Delta} \, 
(\mathbf{T}_{\text{A,C}}^{(1)} \, \bm{\Omega}_\text{A})^\trasp \right] 
 \label{eq:Phi0}
\\
\bm{\Phi}_{\text{A},j,\text{C}}[1] & \eqdef
e^{j  2 \pi \nu_{\text{A,C}}} \, \mathbf{D}_{\text{A,C}}^* \, \Rb_{\overline{\yb}_j\overline{\yb}_j^*}^{2 \, \nu_{\text{A,C}}}[1] \, \mathbf{D}_{\text{A,C}}^*
\nonumber \\ & =  
g_{\text{A,C}}^2 \,  e^{\jmath 2 \pi (j-1) \sin(\theta_{\text{A,C}})} 
\nonumber \\ & \hspace{20mm} \cdot 
(\mathbf{T}_{\text{A,C}}^{(1)} \,
\bm{\Omega}_\text{A}) \, \bm{\Delta} \, (\mathbf{T}_{\text{A,C}}^{(0)} \,
\bm{\Omega}_\text{A})^\trasp \:.
\label{eq:Phi+1}
\earr
The proposed estimation algorithm relies on the fact that
\barr
\bm{\Phi}_{\text{A,C}} & \eqdef \sum_{j=1}^J \sum_{r=-1}^{1} \bm{\Phi}_{\text{A},j,\text{C}}[r] 
\nonumber \\ & =
g_{\text{A,C}}^2  \, {\rm D}_J^*\left(\sin(\theta_{\text{A,C}})\right) \, \mathbf{C}_{\text{A,C}} \, \bm{\Omega}_\text{A} \, \bm{\Delta} \, \bm{\Omega}_\text{A}^\trasp \,
\mathbf{C}_{\text{A,C}}^\trasp
\label{eq:Phi}
\earr
for $\text{C} \in \{\text{L, N}\}$, where the Dirichlet function has been defined in \eqref{eq:Dic}, the matrix 
$\mathbf{C}_{\text{A,C}} \eqdef  \mathbf{T}_{\text{A,C}}^{(0)}
+ \mathbf{T}_{\text{A,C}}^{(1)} \in \Rset^{P \times P}$ is circulant 
by construction,
whose first column is given by
\begin{multline}
\mathbf{c}_{\text{A,C}} \eqdef \Big(\psi(-\tau_{\text{A,C}}), \psi(T_\text{c}-\tau_{\text{A,C}}),
\ldots,
 \\
\psi(\Lcp \, T_\text{c}-\tau_{\text{A,C}}), 0, \ldots, 0 \Big)^\trasp \: .
\end{multline}
Such a circulant matrix can be easily diagonalized as
\be
\mathbf{C}_{\text{A,C}}=\bm{W}_P \,  \diag(\vb_{\text{A,C}}) \,
\bm{W}_P^\herm
\label{eq:CC}
\ee
where the $p$th entry of
\be
\vb_{\text{A,C}} \eqdef \sqrt{P} \,\bm{W}_P^\herm \, \mathbf{c}_{\text{A,C}} \in \Cset^P
\ee
is given by $\{\vb_{\text{A,C}}\}_p
= \Psi_{\text{A,C}}[p] \, e^{-\jmath \frac{2 \pi}{T_\text{s}} \tau_{\text{A,C}} \,p}$,  with  
\be
\Psi_{\text{A,C}}[p]  \approx 
\begin{cases}
\frac{1}{T_\text{c}} \, \Psi\left(\frac{p}{T_\text{s}}\right) \:,
\\  \hspace{10mm} \text{for $p \in \left\{0, 1, \ldots, \frac{P}{2}-1\right\}$}
\\
\frac{1}{T_\text{c}} \, \Psi\left(\frac{p- P}{T_\text{s}}\right) \, e^{\jmath \frac{2 \pi}{T_\text{c}} \chi_{\text{A,C}}} \:,
\\  \hspace{10mm} \text{for $p \in \left\{\frac{P}{2}-1, \frac{P}{2}, \ldots, P-1\right\}$} 
\end{cases}
\label{eq:59}
\ee
and $\Psi(f) \approx 0$ for $f \not \in (-0.5/T_\text{c},0.5/T_\text{c})$
being the Fourier transform of $\psi(t)$.
Is is worth noticing that
$\Psi_{\text{A,C}}[0], \Psi_{\text{A,C}}[1], \ldots, \Psi_{\text{A,C}}[\frac{P}{2}-1]$ are equal to the corresponding coefficients of the
$P$-point DFT of $\psi(\ell \, T_\text{c})$ and, thus, they are known at the BS.

To acquire the delay $\tau_{\text{A,C}}$,
we observe that, by substituting \eqref{eq:CC} in \eqref{eq:Phi},
it results that
\begin{multline}
\bm{W}_P^\herm \, \bm{\Phi}_{\text{A,C}} \, \bm{W}_P^*=
g_{\text{A,C}}^2 \, {\rm D}_J^*\left(\sin(\theta_{\text{A,C}})\right)
\\ \cdot  
\Eb_{\text{A,C}} \, \bm{\Psi}_{\text{A,C}} \,
\bm{\Upsilon}  \, \bm{\Psi}_{\text{A,C}} \, \Eb_{\text{A,C}}
\end{multline}
for $\text{C} \in \{\text{L, N}\}$, where we have defined the matrices
\barr
\Eb_{\text{A,C}} & \eqdef \diag\left(1, e^{-\jmath \frac{2 \pi}{T_\text{s}} \tau_{\text{A,C}}}, \ldots,
e^{-\jmath \frac{2 \pi}{T_\text{s}} \tau_{\text{A,C}} (P-1)} \right)
\\
\bm{\Psi}_{\text{A,C}} & \eqdef \diag\left(\Psi_{\text{A,C}}[0],
\Psi_{\text{A,C}}[1], \ldots, \Psi_{\text{A,C}}[P-1]\right)
\\
\bm{\Upsilon} & \eqdef \bm{W}_P^\herm \, \bm{\Omega}_\text{A} \, 
\bm{\Delta} \, \bm{\Omega}_\text{A}^\trasp \,
\bm{W}_P^*  \in \Cset^{P \times P} \:.
\earr
For $p \in  \{0,1,\ldots, P-1\}$, it is readily seen that
the $p$th diagonal entry of $\bm{W}_P^\herm \, \bm{\Phi}_{\text{A,C}} \, \bm{W}_P^*$
is given by
\begin{multline}
\{\bm{W}_P^\herm \, \bm{\Phi}_{\text{A,C}} \, \bm{W}_P^*\}_{p,p} \\ =
g_{\text{A,C}}^2 \, {\rm D}_J^*\left(\sin(\theta_{\text{A,C}})\right)  
\Psi_{\text{A,C}}^2[p] \, \{\bm{\Upsilon}\}_{p,p}
\, e^{-j \frac{4 \pi}{T_\text{s}} \tau_{\text{A,C}} \,p} \: .
\end{multline}

At this point, let us introduce 
the {\em one-dimensional} cost function defined in
\eqref{eq:I} at the top of the next page. 
\begin{figure*}[!t]
\normalsize
\barr
\mathcal{I}_{\text{A,C}}(\beta) & \eqdef
\left| \sum_{p=0}^{P/2-1} \{\bm{W}_P^\herm \, \bm{\Phi}_{\text{A,C}} \, \bm{W}_P^*\}_{p,p}
\, (\Psi_{\text{A,C}}^2[p])^*  \, \{\bm{\Upsilon}\}_{p,p}^* \, e^{j \frac{4 \pi}{T_\text{s}} \beta \, p}
\right|
\nonumber \\ & = |g_{\text{A,C}}|^2 \, \left|{\rm D}_J^*\left(\sin(\theta_{\text{A,C}})\right) \right| \, 
\left| \sum_{p=0}^{P/2-1} \left|\Psi_{\text{A,C}}[p]\right|^4
\, \left|\{\bm{\Upsilon}\}_{p,p}\right|^2 e^{-j \frac{4 \pi}{T} (\tau_{\text{A,C}}-\beta) \, p}
\right|
\:, \quad \text{with $\beta \in [0, \Delta_\text{max}]$}
\label{eq:I}
\earr
\hrulefill
\end{figure*}
where $\Delta_\text{max} \ge \max\{\tau_{\text{A,L}},\tau_{\text{A,N}}\}$
is a known upper bound on the maximum delay of the AU channel.
By resorting to the triangle inequality, it can be verified that, for $\text{C} \in \{\text{L, N}\}$,  
$\mathcal{I}_{\text{A,C}}(\beta)$ takes its maximum value when 
$\beta=\tau_{\text{A,C}}+i\, T_\text{s}/2$ ($i \in \Zset$). 
Hence, under the assumption $\Delta_\text{max} < T_\text{s}/2$,
acquisition of 
$\tau_{\text{A,C}}$ can be pursued by  searching
for the global maximum of \eqref{eq:I}
over the interval $[0, \Delta_\text{max}] \subset [0, T_\text{s}/2)$.

In practice, an estimate $\widehat{\tau}_{\text{A,C}}$ of the
delay $\tau_{\text{A,C}}$ is derived by searching for the peak of
the finite-sample version $\widehat{\mathcal{I}}_{\text{A,C}}(\beta)$ of the cost function
$\mathcal{I}_{\text{A,C}}(\beta)$ defined in \eqref{eq:I}, where
$\widehat{\mathcal{I}}_{\text{A,C}}(\beta)$ is obtained by replacing
$\Rb_{\overline{\yb}_j\overline{\yb}_j^*}^{2 \, \nu_{\text{A,C}}}[r]$
in \eqref{eq:Phi-1}--\eqref{eq:Phi+1},
for $r \in \{-1,0,1\}$,
with their corresponding estimates 
$\widehat{\bm R}_{\overline{\yb}_j\overline{\yb}_j^*}^{2 \, \widehat{\nu}_{\text{A,C}}}[r]$
[see \eqref{eq:Rciclest}],
evaluated at the
estimated cycle frequency
$2\, \widehat{\nu}_{\text{A,C}}$ acquired through the 
algorithm in Subsection~II-A-1.

\subsection{Pilot-aided 
estimation of complex gain and AoA parameters}
\label{sec:gain}

After acquiring the Doppler and delay parameters blindly, 
we propose to estimate the complex gain and
AoA unknowns of the AU, by capitalizing on its pilots 
transmitted in the training intervals 
$\mathcal{N}_\text{A}=\{n_{\text{A},1}, n_{\text{A},2},  \ldots, n_{\text{A},N_\text{A,train}}\}$.  
For the present, Doppler and delay parameters
are assumed to be perfectly known.
The starting point of 
the pilot-aided estimation step 
is 
\eqref{eq:y_j}, which collects the 
signal received by the $j$-th antenna
after CP removal and DFT. 

For $n \in \mathcal{N}_\text{A}$
and $j \in \{1,2,\ldots, J\}$, 
it is convenient to rewrite 
\eqref{eq:y_j} as follows
\begin{multline}
\yb_j[n] = 
g_{\text{A,L}} \, \eta_{\text{A,L}}^{j-1}   
\, \mathbf{p}_{\text{A,L}}[n] 
\\ + 
g_{\text{A,N}} \, \eta_{\text{A,N}}^{j-1}   
\, \mathbf{p}_{\text{A,N}}[n] +
\dw_{\text{A},j}[n]
\label{eq:y_jj}
\end{multline}
where 
$\mathbf{p}_{\text{A,C}}[n] \eqdef \bm{W}_M^\herm \, \bm{R}_\text{cp} \, 
\mathbf{D}_{\text{A,C}} \,
\mathbf{T}_{\text{A,C}}^{(0)} \, \bm{\Omega}_\text{A}\, \bab[n] \, 
e^{\jmath \, 2 \pi \nu_{\text{A,C}} n}$ is a {\em known} vector, 
$\eta_{\text{A,C}} \eqdef e^{\jmath \pi \sin(\theta_{\text{A,C}})}$ is a 
complex parameter  to be estimated together with $g_{\text{A,C}}$,
for $\text{C} \in \{\text{L, N}\}$, 
whereas  
$\dw_{\text{A},j}[n] \eqdef \Mb_{\text{T},j} \, \bsb[n]  + \wb_j[n]$
represents disturbance from the AU viewpoint.

Given the AU symbol vector $\bab[n]$, the random vector $\yb_j[n] $ is circular,
for $j \in \{1,2,\ldots, J\}$.
By gathering all the training blocks received by the $j$-th antenna in 
the matrix $\Yb_{\text{A},j} \eqdef (\yb_j[n_{\text{A},1}], \yb_j[n_{\text{A},2}], \ldots, \yb_j[n_{\text{A},N_\text{A,train}}]) \in \Cset^{M \times N_\text{A,train}}$, one gets the compact matrix model
\be
\Yb_{\text{A},j} = 
g_{\text{A,L}} \,  \eta_{\text{A,L}}^{j-1} 
\, \mathbf{P}_{\text{A,L}}+  g_{\text{A,N}} \,  \eta_{\text{A,N}}^{j-1} 
\, \mathbf{P}_{\text{A,N}}+ \Db_{\text{A},j}
\label{eq:Y_j}
\ee
with 
\barr
\mathbf{P}_{\text{A,C}}  & \eqdef (\mathbf{p}_{\text{A,C}}[n_{\text{A},1}], 
\mathbf{p}_{\text{A,C}}[n_{\text{A},2}], 
\nonumber \\ 
& \hspace{20mm}  \ldots, 
\mathbf{p}_{\text{A,C}}[n_{\text{A},N_\text{A,train}}]) \in \Cset^{M \times N_\text{A,train}}
\\
\Db_{\text{A},j}  & \eqdef (\dw_{\text{A},j}[n_{\text{A},1}], \dw_{\text{A},j}[n_{\text{A},2}], 
\nonumber \\
& \hspace{20mm}  \ldots, 
\dw_{\text{A},j}[n_{\text{A},N_\text{A,train}}]) \in \Cset^{M \times N_\text{A,train}} 
\earr
for $\text{C} \in \{\text{L, N}\}$ and $j \in \{1,2,\ldots, J\}$.
From now on, it is assumed that  
$\eta_{\text{A,L}}$ is distinct from $\eta_{\text{A,N}}$.

Since the correlation properties of $\Db_{\text{A},j}$
are unknown due to the multiple access interference arising from
the transmission of the TU, estimation strategies involving 
the second-order statistics of the disturbance,
such as best linear unbiased (BLU) or linear MMSE estimators, 
cannot be implemented in this case. 
Therefore, we resort to the {\em least-squares  (LS)} criterion \cite{Kay} 
to acquire channel gains and AoAs of the AU, whose salient 
feature is that no probability assumptions are made about the received data.

The four parameters $\eta_{\text{A,L}}$, $g_{\text{A,L}}$, 
$\eta_{\text{A,N}}$, and $g_{\text{A,N}}$ have to be estimated.
The LS estimator is defined as
\be
(\widehat{g}_{\text{A,L}}, \widehat{g}_{\text{A,N}},
\widehat{\eta}_{\text{A,L}}, \widehat{\eta}_{\text{A,N}})  = 
\arg \min_{\shortstack{\footnotesize
$\rho_\text{L}, \rho_\text{N} \in \Cset$ \\ \footnotesize 
$|\varrho_\text{L}|=|\varrho_\text{N}|=1$}} 
{\mathcal L}_A(\rho_\text{L}, \rho_\text{N},\varrho_\text{L},\varrho_\text{N})
\label{eq:LS-gain}
\ee
with 
\be
{\mathcal L}_A(\rho_\text{L}, \rho_\text{N},\varrho_\text{L},\varrho_\text{N}) \eqdef 
\sum_{j=1}^J \Big \|  \Yb_{\text{A},j} - \sum_{\text{C} \in \{\text{L, N}\}}
\rho_\text{C} \,    
\, \varrho_{\text{C}}^{j-1}
\, \mathbf{P}_{\text{A,C}} 
\Big \|^2 \:.
\label{eq:L}
\ee
Significant computational savings follow from the observation
that, for any {\em given} $\eta_{\text{A,L}}$
and $\eta_{\text{A,N}}$, 
the problem of finding
the LS estimates of 
$g_{\text{A,L}}$ and $g_{\text{A,N}}$, 
corresponding to that $\eta_{\text{A,L}}$
and $\eta_{\text{A,N}}$, are obtained by 
solving the $ 2 \times 2$ linear system:
\barr
\frac{\partial}{\partial \rho_\text{L}^*} \, 
{\mathcal L}_A(\rho_\text{L}, \rho_\text{N},\varrho_\text{L},\varrho_\text{N}) & =
- \Lambda_{\text{L},0}+ \Lambda_{\text{L},1} \, \rho_\text{L} 
+ \Lambda_{\text{L},2} \, \rho_\text{N}=0 
\label{eq:sys-1}
\\
\frac{\partial}{\partial \rho_\text{N}^*} \, 
{\mathcal L}_A(\rho_\text{L}, \rho_\text{N},\varrho_\text{L},\varrho_\text{N}) & =
- \Lambda_{\text{N},0}+ \Lambda_{\text{N},1} \, \rho_\text{N} 
+ \Lambda_{\text{N},2} \, \rho_\text{L}=0 
\label{eq:sys-2}
\earr
with
\barr
\Lambda_{\text{C},0} 
& \eqdef \sum_{j=1}^J \trace\left(\Yb_{\text{A},j} \, \mathbf{P}_{\text{A,C}}^\herm \,
\, \eta_{\text{A,C}}^{-(j-1)}\right)
\\ 
\Lambda_{\text{C},1} & \eqdef J \, \trace(\mathbf{P}_{\text{A,C}} \, 
\mathbf{P}_{\text{A,C}}^\herm)
= J \, \| \mathbf{P}_{\text{A,C}} \|^2
\\
\Lambda_{\text{C},2} & \eqdef \sum_{j=1}^J \trace\left(\mathbf{P}_{\text{A},\overline{\text{C}}} 
\, \mathbf{P}_{\text{A,C}}^\herm \,
(\eta_{\text{A},\overline{\text{C}}} \,  \eta_{\text{A,C}}^*)^{j-1}\right)
\earr
where $\overline{\text{C}}=\text{N}$ when $\text{C}=\text{L}$, whereas $\overline{\text{C}}=\text{L}$  when $\text{C}=\text{N}$.
The solution of the system \eqref{eq:sys-1}-\eqref{eq:sys-2} is given by
\barr
\widehat{g}_{\text{A,L}} & = \frac{\Lambda_{\text{L},0} \, \Lambda_{\text{N},1}-
\Lambda_{\text{L},2} \, \Lambda_{\text{N},0}}{\Lambda_{\text{L},1} \, \Lambda_{\text{N},1}-\Lambda_{\text{L},2} \, \Lambda_{\text{N},2}}
\label{eq:ga0}
\\
\widehat{g}_{\text{A,N}} & =
\frac{\Lambda_{\text{L},1} \, \Lambda_{\text{N},0}-\Lambda_{\text{L},0} \, 
\Lambda_{\text{N},2}}
{\Lambda_{\text{L},1} \, \Lambda_{\text{N},1}-\Lambda_{\text{L},2} \, \Lambda_{\text{N},2}} \: .
\label{eq:ga1}
\earr
By substituting \eqref{eq:ga0}-\eqref{eq:ga1}
into \eqref{eq:LS-gain},  the LS estimates of $\eta_{\text{A,L}}$
and $\eta_{\text{A,N}}$ can be found by
\be
(\widehat{\eta}_{\text{A,L}}, \widehat{\eta}_{\text{A,N}})  = 
\arg \min_{|\varrho_\text{L}|=|\varrho_\text{N}|=1} 
{\mathcal L}_A(\widehat{g}_{\text{A,L}}, \widehat{g}_{\text{A,N}},\varrho_\text{L},\varrho_\text{C})
\label{eq:LS-angle} \: .
\ee
In general, the cost function in \eqref{eq:LS-angle} is multimodal.
However, there exist computationally-efficient global optimization 
procedures, which try to provide an efficiently global
optimization by partition of definition space at multi-scale levels, see, e.g., \cite{Sun,Opt-book}.
Such methods have a stable convergence speed and the optimization solution
is independent of the selection of initial solutions \cite{Sun}. 

In practice,  the previously obtained estimates of the Doppler
and delay parameters (see Subsection~III-A-1 and III-A-2)
have to be used for LS acquisition
of the complex gain and AoA unknowns of the AU. 
Specifically,  the LS problem to be solved comes from replacing 
$\mathbf{P}_{\text{A,C}} $ in \eqref{eq:L} with its 
corresponding estimate 
$\widehat{\mathbf{P}}_{\text{A,C}} \eqdef (\widehat{\mathbf{p}}_{\text{A,C}}[n_{\text{A},1}], 
\widehat{\mathbf{p}}_{\text{A,C}}[n_{\text{A},2}], \ldots, 
\widehat{\mathbf{p}}_{\text{A,C}}[n_{\text{A},N_\text{A,train}}])$, where
$\widehat{\mathbf{p}}_{\text{A,C}}[n] \eqdef \bm{W}_M^\herm \, \bm{R}_\text{cp} \, 
\widehat{\mathbf{D}}_{\text{A,C}} \,
\widehat{\mathbf{T}}_{\text{A,C}}^{(0)} \, \bm{\Omega}_\text{A} \, \bab[n] 
\, e^{\jmath \, 2 \pi \widehat{\nu}_{\text{A,C}} n}$, for any $n \in \mathcal{N}_\text{A}$, 
and $\widehat{\mathbf{T}}_{\text{A,C}}^{(0)}$ is built from 
\eqref{eq:T0} by using $\widehat{\tau}_{\text{A,C}}$ 
in lieu of $\tau_{\text{A,C}}$.

\subsection{Computational issues}
With reference to Fig.~\ref{fig:fig_sum_alg}, the computational 
complexity of the overall 
estimation algorithm for the AU is mainly dominated by 
the first block devoted to the Doppler shifts acquisition, i.e., the CCCM 
estimation. In particular, the CCCM estimation entails 
$\mathcal{O}(N_0^2 \, P^2 \, J)$ floating point operations (flops) and, thus,
the complexity burden quadratically grows with the length $N_0$ of the 
observation interval.

\subsection{NOMA is used by TUs in the uplink}
\label{sec:NOMA-TU-A}
So far, we have assumed that the TUs transmit to the BS using 
the OMA scheme. We now discuss how the proposed 
channel estimation algorithm for the AU will work 
if nonorthogonal RBs are assigned to TUs in the uplink. 

In this scenario, a single AU may send noncircular symbols using the same RBs 
shared by a group of static TUs that concurrently transmit to the BS by employing 
a power-domain NOMA scheme with circular modulation formats. 
In such a case, the proposed blind estimation procedure of Doppler and delay parameters 
is unaffected, since the terrestrial transmissions
do not contribute to the CCCM \eqref{eq:Ralpha} due to the modulation and 
Doppler diversities. 
On the other hand, the uplink signals of the TUs
appear in the disturbance term  $\dw_{\text{A},j}[n]$ in \eqref{eq:y_jj},
thus impacting on the pilot-aided estimation of complex gain and 
AoA parameters. 
However, existing schemes
in the power-domain NOMA literature (e.g., \cite{Ding.2016,Cheng.2018})
creates multiple groups, each with one ``weak" TU at the cell edge and one 
``strong" TU  at the cell center. For such a NOMA scheme, the power of  
$\dw_{\text{A},j}[n]$ will be mainly dictated by the strong TU and, hence,  
compared with the single TU case considered in this paper, 
the performance degradation of the proposed LS estimator \eqref{eq:LS-gain} will be  
negligible. 

\section{Channel estimation for the terrestrial user}
\label{sec:estimation-TU}

The estimation of the AU channel follows
an array signal processing approach, where 
the parameters characterizing
the AU channel are estimated. 
We resort to a completely different approach for
the estimation of the TU channel.
More precisely,  
the TU channel estimator passes to the 
SIC detector the diagonal time-invariant matrix 
$\bm{H}_{\text{T},j}\equiv \bm{M}_{\text{T},j}$,
for each antenna index $j \in \{1,2,\ldots, J\}$.
Since the TU channel is time invariant, we pursue 
a pilot-aided approach for estimating $\bm{M}_{\text{T},j}$,
which relies on the $Q_\text{T}$ pilot symbols transmitted
by the TU at the subcarriers
$\mathcal{M}_\text{T} =
\{m_{\text{T},0}, m_{\text{T},1}, 
\ldots, m_{\text{T},Q_\text{T}-1}\}$  during 
the $N_\text{T,train}$ 
symbol intervals $\mathcal{N}_\text{T}=\{n_{\text{T},1}, n_{\text{T},2},  \ldots, n_{\text{T}
,N_\text{T,train}}\}$.  
This task
is developed by assuming perfect 
knowledge of the AU channel matrices $\bm{H}_{\text{A},1}[n], \bm{H}_{\text{A},2}[n],\ldots,
\bm{H}_{\text{A},J}[n]$, which can be acquired as proposed in 
Section~\ref{sec:estimation-AU}.
To this aim, according to \eqref{eq:mu-TU},
instead of separately estimating 
the parameters (i.e., time delays, AoAs, and channel gains) characterizing the TU channel, 
we choose to directly estimate the vector 
\be
\gb_{\text{T},j} \eqdef \sum_{k=1}^{K_\text{T}} \gb_{\text{T},j,k} \in \Cset^{\Lcp}
\ee
which gathers all the aggregated state information of the TU channel
corresponding to the $j$th receiving antenna.

Let us consider the frequency-domain received data given by \eqref{eq:y_j}.
Denoting by 
\barr
\yb_{\text{T},j}[n] 
& \eqdef (y_j^{[\Lcp+m_{\text{T},0}]}[n], y_j^{[\Lcp+m_{\text{T},1}]}[n],
\nonumber \\ & \hspace{20mm} \ldots,
y_j^{[\Lcp+m_{\text{T},Q_\text{T}-1}]}[n])^\trasp \in \Cset^{Q_\text{T}}
\\
\wb_{\text{T},j}[n] 
& \eqdef (w_j^{[\Lcp+m_{\text{T},0}]}[n], w_j^{[\Lcp+m_{\text{T},1}]}[n],
\nonumber \\ & \hspace{20mm} \ldots,
w_j^{[\Lcp+m_{\text{T},Q_\text{T}-1}]}[n])^\trasp \in \Cset^{Q_\text{T}}
\earr
the vectors
containing the entries of 
 $\yb_j[n]$ and $\wb_j[n]$ at the TU pilot locations,
respectively, we get
\be
\yb_{\text{T},j}[n] = 
\mathbf{P}_{\text{T}}[n] \, \gb_{\text{T},j}  + \dw_{\text{T},j}[n]
\label{eq:y_Tj}
\ee
for $n \in \mathcal{N}_\text{T}$
and $j \in \{1,2,\ldots, J\}$, where 
\barr
\mathbf{P}_{\text{T}}[n] \eqdef \sqrt{M} \, \bm{S}_\text{T}([n]) \, \bm{W}_M^\herm \, \Lb 
\in \Cset^{Q_\text{T} \times \Lcp}
\earr
is a {\em known} matrix, 
with the pilot matrix $\bm{S}_\text{T}([n]) \in \Cset^{Q_\text{T} \times M}$ obtained by picking up 
the  rows of $\diag(\bsb[n])$ at positions
$\{m_{\text{T},0}+1, m_{\text{T},1}+1, 
\ldots, m_{\text{T},Q_\text{T}-1}+1\}$ for $n \in \mathcal{N}_\text{T}$, 
whereas  
$\dw_{\text{T},j}[n] \eqdef \Mb_{\text{A},j}[n] \, \bab[n]  + \wb_j[n]$
represents disturbance from the TU viewpoint, with 
$\Mb_{\text{A},j}[n] \in \Cset^{Q_\text{T} \times M}$
obtained by taking
the  rows of the known matrix $\bm{H}_{\text{A},j}[n]$ corresponding to the indexes $\{m_{\text{T},0}+1, m_{\text{T},1}+1, 
\ldots, m_{\text{T},Q_\text{T}-1}+1\}$.
The training blocks received by the $j$-th antenna for $n \in \mathcal{N}_\text{T}$ are 
gathered into the $(Q_\text{T} N_\text{T,train})$-dimensional complex vector 
\be
\yb_{\text{T},j} \eqdef (\yb_{\text{T},j}^\trasp[n_{\text{T},1}], 
\yb_{\text{T},j}^\trasp[n_{\text{T},2}], \ldots, \yb_{\text{T},j}^\trasp[n_{\text{T},N_\text{T,train}}])^\trasp 
\ee
thus yielding, for $j \in \{1,2,\ldots, J\}$,  
\be
\yb_{\text{T},j} = 
\mathbf{P}_{\text{T}} \, \gb_{\text{T},j}  + \dw_{\text{T},j}
\label{eq:y_T}
\ee
where 
\barr
\mathbf{P}_{\text{T}} & \eqdef (\mathbf{P}_{\text{T}}^\trasp[n_{\text{T},1}], 
\mathbf{P}_{\text{T}}^\trasp[n_{\text{T},2}], 
\nonumber \\ & \hspace{20mm} \ldots
\mathbf{P}_{\text{T}}^\trasp[n_{\text{T},N_\text{T,train}}])^\trasp \in 
\Cset^{(Q_\text{T} N_\text{T,train}) \times \Lcp}
\\
\dw_{\text{T},j} & \eqdef (\dw_{\text{T},j}^\trasp[n_{\text{T},1}], 
\dw_{\text{T},j}^\trasp[n_{\text{T},2}], 
\nonumber \\ & \hspace{20mm} 
\ldots, \dw_{\text{T},j}^\trasp[n_{\text{T},N_\text{T,train}}])^\trasp \in \Cset^{Q_\text{T} N_\text{T,train}} \:.
\earr
Finally, we collect the data received by all the antennas of the BS into 
$\yb_\text{T} \eqdef (\yb_{\text{T},1}^\trasp, \yb_{\text{T},2}^\trasp, \ldots, \yb_{\text{T},J}^\trasp)^\trasp
\in \Cset^{J Q_\text{T} N_\text{T,train}}$, thus  yielding 
the vector signal model
\be
\yb_\text{T} = \left(\I_{J} \otimes \mathbf{P}_{\text{T}}\right)  \gb_{\text{T}} + \dw_\text{T} 
\label{eq:y_TT}
\ee
where 
\barr
\gb_{\text{T}} & \eqdef (\gb_{\text{T},1}^\trasp, \gb_{\text{T},2}^\trasp, \ldots, 
\gb_{\text{T},J}^\trasp)^\trasp \in \Cset^{J \Lcp}
\\ 
\dw_\text{T} & \eqdef (\dw_{\text{T},1}^\trasp, \dw_{\text{T},2}^\trasp, \ldots, \dw_{\text{T},J}^\trasp)^\trasp
\in \Cset^{J Q_\text{T} N_\text{T,train}} \;.
\earr
We assume that $\mathbf{P}_{\text{T}}$ is full-column rank, i.e., 
$\rank(\mathbf{P}_{\text{T}})=\Lcp$  
and $Q_\text{T} N_\text{T,train} \ge \Lcp$, 
which ensures that 
$\gb_{\text{T}}$  is uniquely identifiable
from \eqref{eq:y_TT} in the absence of the disturbance $\dw_{\text{T}}$. 

There are different estimation strategies that can be pursued to estimate
the TU channel from \eqref{eq:y_TT}.
We focus on estimators belonging 
to the family of the classical estimation approaches \cite{Kay}, for which 
$\gb_{\text{T}}$ is viewed as deterministic but unknown vector. 
The simplest estimator is represented by the LS one (see, e.g., \cite{Kay}):
\barr
\widehat{\gb}_{\text{T}, \text{LS}} & \eqdef \arg \min_{\gb_{\text{T}} \in \Cset^{J \Lcp}} \|\yb_\text{T} 
- \mathbf{P}_{\text{T}} \, \gb_{\text{T}}\|^2 
\nonumber \\ &
= \left[\I_{J} \otimes\left(\mathbf{P}_{\text{T}}^\herm \, \mathbf{P}_{\text{T}}\right)^{-1} \mathbf{P}_{\text{T}}^\herm \right] \yb_\text{T} 
\label{eq:LS}
\earr
which does not exploit the CSI knowledge of the AU and, thus, performs poorly in a NOMA setting
(see Section~\ref{sec:simul}). On the other hand, the BS can jointly exploits
the knowledge of the AU channel matrices $\bm{H}_{\text{A},1}[n], \bm{H}_{\text{A},2}[n],\ldots,
\bm{H}_{\text{A},J}[n]$ and the noncircularity of the AU symbols, 
by resorting to the {\em best  WL unbiased (BWLU)} estimator, which 
is restricted to be WL in the data, it is unbiased, and it has minimum variance.
In matrix form, a WL estimator is defined by
\barr
\widehat{\gb}_{\text{T}} & = \Ab_1 \, \yb_\text{T} + \Ab_2 \, \yb_\text{T}^*
\nonumber \\ &
= 
\underbrace{(\Ab_1, \Ab_2)}_{\Ab \in \Cset^{(J \Lcp) \times (2 J Q_\text{T} N_\text{T,train})}} \, \underbrace{\begin{pmatrix}
\yb_\text{T} \\
\yb_\text{T}^*
\end{pmatrix}}_{\ybtilde_\text{T} \in \Cset^{2 J Q_\text{T} N_\text{T,train}} } =
\Ab \, \ybtilde_\text{T}
\earr
where $\Ab_1, \Ab_2 \in \Cset^{(J \Lcp) \times (J Q_\text{T} N_\text{T,train})}$  and 
\be
\ybtilde_\text{T} = \widetilde{\mathbf{P}}_{\text{T}} \, \gb_{\text{T}} +
\Jb_{J Q_\text{T} N_\text{T,train}} \, \widetilde{\mathbf{P}}_{\text{T}}^* \, \gb_{\text{T}}^*+ \dwtilde_\text{T}
\label{eq:yTtilde}
\ee
with 
\be
\widetilde{\mathbf{P}}_{\text{T}}  \eqdef (\I_J \otimes \mathbf{P}_{\text{T}}^\trasp, 
\bm{O}_{(J Q_\text{T} N_\text{T,train}) \times (J \Lcp)}^\trasp)^\trasp
\in  \Cset^{(2 J Q_\text{T} N_\text{T,train}) \times (J \Lcp)}
\ee
$\Jb_{J Q_\text{T} N_\text{T,train}}  \in \Rset^{(2 J Q_\text{T} N_\text{T,train}) 
\times (2 J Q_\text{T} N_\text{T,train})}$
defined in \eqref{eq:Jmat}, and, finally,  
$\dwtilde_\text{T} \eqdef (\dw_\text{T}^\trasp, \dw_\text{T}^\herm)^\trasp
\in \Cset^{2 J Q_\text{T} N_\text{T,train}}$.
It is readily verified that 
$\dwtilde_\text{T}$ has zero mean, i.e., 
$\Es(\dwtilde_\text{T})=\bm{0}_{2 J Q_\text{T} N_\text{T,train}}$,  
and 
\be
\Rb_{\dwtilde_\text{T}\dwtilde_\text{T}}
\eqdef \Es(\dwtilde_\text{T} \,
\dwtilde_\text{T}^\herm) = \Mbtilde_{\text{A}} \, \Mbtilde_{\text{A}}^\herm  + \sigma_w^2 \, \I_{2 J Q_\text{T} N_\text{T,train}}
\ee 
where 
\be
\Mbtilde_{\text{A}} \eqdef 
\begin{pmatrix} \Mb_{\text{A}} \\
\Mb_{\text{A}}^* \, [\I_{N_\text{T,train}} \otimes \mathbf{\Delta}] \end{pmatrix} \in 
\Cset^{(2 J Q_\text{T} N_\text{T,train}) \times 
(M N_\text{T,train})}
\ee
with 
\barr
\Mb_{\text{A}}  & \eqdef (\Mb_{\text{A},1}^\trasp, \Mb_{\text{A},2}^\trasp, \ldots, 
\Mb_{\text{A},J}^\trasp)^\trasp \in \Cset^{(J Q_\text{T} N_\text{T,train}) \times 
(M N_\text{T,train})}
\nonumber \\
\Mb_{\text{A},j}  & \eqdef \diag(\Mb_{\text{A},j}[n_{\text{T},1}], 
\Mb_{\text{A},j}[n_{\text{T},2}], \ldots, 
\Mb_{\text{A},j}[n_{\text{T},N_\text{T,train}}]) \: .
\earr
The key point of the proposed BWLU estimation approach is the 
fact that the correlation matrix $\Rb_{\dwtilde_\text{T}\dwtilde_\text{T}}$ is 
known at the BS after acquiring the AU channel state 
(see Section~\ref{sec:estimation-AU}).

According to \eqref{eq:yTtilde}, 
in order for $\widehat{\gb}_{\text{T}}$ to be unbiased, we require
$\Es(\widehat{\gb}_{\text{T}})= \Ab \, \Es(\ybtilde_\text{T})  = \gb_{\text{T}}$,
which is tantamount to imposing the unbiased constraint 
\begin{multline}
\Ab 
\underbrace{\begin{pmatrix} 
\I_J \otimes \mathbf{P}_{\text{T}} & 
\mathbf{O}_{(J Q_\text{T} N_\text{T,train}) \times (J \Lcp)}
\\
\mathbf{O}_{(J Q_\text{T} N_\text{T,train}) \times (J \Lcp)} & 
\I_J \otimes \mathbf{P}_{\text{T}}^*
\end{pmatrix}}_{\bm{\Pi}_\text{T} \in \Cset^{(2 J Q_\text{T} N_\text{T,train}) \times 
(2 J \Lcp)}} \\ =
\underbrace{\left(\I_{J \Lcp}, \mathbf{O}_{(J \Lcp) 
\times (J \Lcp)}\right)}_{\bm{\Theta} \in \Rset^{(J \Lcp)  \times (2 J \Lcp) }
}
\label{eq:construn}
\end{multline}
The BWLU estimator is found by minimizing the variance at the estimator output
\barr
\text{Var}(\widehat{\gb}_{\text{T}}) & \eqdef \Es\left(\left\|\widehat{\gb}_{\text{T}}-\gb_{\text{T}}\right\|^2\right) 
 \nonumber \\ & =   
\trace\left[\Ab \, \Es(\ybtilde_\text{T} \, \ybtilde_\text{T}^H) \, \Ab^\herm \right] -
\|\gb_{\text{T}}\|^2
 \nonumber \\ & =   
\trace(\Ab \, \Rb_{\dwtilde_\text{T}\dwtilde_\text{T}} \, \Ab^\herm)
\label{eq:variance}
\earr
where we have also taken into account the constraint \eqref{eq:construn}.
The classical approach for minimizing \eqref{eq:variance} subject to 
the unbiased constraint \eqref{eq:construn} is the method of Lagrange multipliers,
which yields the solution (see, e.g., \cite{Kay})
\be
\widehat{\gb}_{\text{T}, \text{BWLU}} = \bm{\Theta} 
\left(\bm{\Pi}_\text{T}^\herm \, \Rb_{\dwtilde_\text{T}\dwtilde_\text{T}}^{-1} 
\bm{\Pi}_\text{T}\right)^{-1} \bm{\Pi}_\text{T}^\herm \, \Rb_{\dwtilde_\text{T}\dwtilde_\text{T}}^{-1}
\ybtilde_\text{T}
\label{eq:BWLU}
\ee
whose corresponding minimum variance is given by
\be
\text{Var}(\widehat{\gb}_{\text{T}, \text{BWLU}}) 
= \trace \left[\bm{\Theta} \left(\bm{\Pi}_\text{T}^\herm \, \Rb_{\dwtilde_\text{T}\dwtilde_\text{T}}^{-1} 
\bm{\Pi}_\text{T}\right)^{-1} \bm{\Theta}^\herm \right] \: . 
\ee
It can be verified by direct inspection that 
\barr
\Ab_{\text{BWLU}} & = \bm{\Theta} 
\left(\bm{\Pi}_\text{T}^\herm \, \Rb_{\dwtilde_\text{T}\dwtilde_\text{T}}^{-1} 
\bm{\Pi}_\text{T}\right)^{-1} \bm{\Pi}_\text{T}^\herm \, \Rb_{\dwtilde_\text{T}\dwtilde_\text{T}}^{-1}
\nonumber \\ &
= (\Ab_{\text{BWLU},1}, \Ab_{\text{BWLU},2})
\earr
with $\Ab_{\text{BWLU},2} \neq \mathbf{O}_{(J \Lcp) \times (J Q_\text{T} N_\text{T,train})}$:
in this case, the BWLU estimator is expected to outperform its BLU counterpart
in the minimum-variance sense.

In practice,  the BWLU estimator is built by using estimates 
$\widehat{\bm{H}}_{\text{A},1}[n], \widehat{\bm{H}}_{\text{A},2}[n],\ldots,
\widehat{\bm{H}}_{\text{A},J}[n]$
of the 
AU channel matrices $\bm{H}_{\text{A},1}[n], \bm{H}_{\text{A},2}[n],\ldots,
\bm{H}_{\text{A},J}[n]$, which are obtained through the three-step approach 
proposed in Section~\ref{sec:estimation-AU}. Such an estimator is 
implemented by replacing $\Rb_{\dwtilde_\text{T}\dwtilde_\text{T}}$ in 
\eqref{eq:BWLU} with its corresponding estimate 
$\widehat{\Rb}_{\dwtilde_\text{T}\dwtilde_\text{T}}$ 
coming from 
$\widehat{\bm{H}}_{\text{A},1}[n], \widehat{\bm{H}}_{\text{A},2}[n],\ldots,
\widehat{\bm{H}}_{\text{A},J}[n]$, for $n \in \mathcal{N}_\text{T}$.

\subsection{Computational issues}
With reference to Fig.~\ref{fig:fig_sum_alg}, the computational complexity of the 
estimation algorithm for the TU is mainly dominated by 
the second block of the channel acquisition for the TU, i.e., the BWLU estimation. In particular, 
the BWLU estimation relies on the inversion of the matrix $\Rb_{\dwtilde_\text{T}\dwtilde_\text{T}}$ which entails 
$\mathcal{O}(J^3 \, Q^3_\text{T} \, N^3_\text{T,train})$ flops 
if one resorts to batch algorithms. Such a matrix inversion 
can also be implemented  by means of a simple and effective recursion, 
similar to the well-known recursive least square algorithm,
with a complexity per iteration of order only 
$\mathcal{O}(J^2 \, Q^2_\text{T} \, N^2_\text{T,train})$ flops. 
Anyway,  the complexity burden is independent of the length $N_0$ of the observation interval.

\subsection{NOMA is used by TUs in the uplink}
\label{sec:NOMA-TU-T}
As in Subsection~III-\ref{sec:NOMA-TU-A}, 
we discuss herein how the proposed 
channel estimation algorithms for the TU will be employed 
when the BS assigns to a single AU the 
same RBs shared by a group of power-domain NOMA TUs.  

If the BS uses orthogonal pilot sequences to distinguish different 
power-domain NOMA TUs \cite{Fan.2019}, then the received signal from a given TU
can be obtained by performing the correlation of 
the received data with the pilot sequence of the desired TU. 
After performing this preprocessing for each user,  the estimators \eqref{eq:LS}
or \eqref{eq:BWLU} will be used with minor modifications to individually 
estimate the channel of each TU. 
On the other hand,  if the BS allocates the same pilot to the power-domain NOMA TUs, the 
estimators \eqref{eq:LS}
or \eqref{eq:BWLU} will estimate a linear combination of the channels
from all NOMA TUs. Such an estimate will still provide a useful description of the combined channel, which could be exploited by the BS to beamform a combination of the data symbols intended for the NOMA TUs \cite{Cheng.2018}.

\section{Numerical results}
\label{sec:simul}

In this section, we provide numerical results aimed at eva\-luating the
performance of the proposed channel estimators for the SG-NOMA scheme.
To this aim, we consider the following simulation setting.
The AU and TU employ OFDM modulation with $M=16$ subcarriers,
CP of length $\Lcp=4$, sampling rate $1/T_\text{c}=625$ kHz.
The signaling format of the AU is 
a $\pi/2$-BPSK.
On the other hand, the TU transmits 
quaternary-phase-shift-keying (QPSK) symbols. 
The carrier frequency is set to $f_\text{carrier}=27$ GHz.
Regarding the training protocol, we 
set $Q_\text{T}=16$ and 
$N_\text{A, train}=N_\text{T, train}=80$.
Unless otherwise specified, 
the TU coherence time is equal to $N_\text{coh}=16.384$ OFDM ksymbols. 

The ULA at the BS has $J=4$ antennas.
The number of paths of the TU-to-BS link is 
fixed to  $K_{\text{T}}=2$.
The Rician factor of the AU channel is  
$K_\text{A}=6$ dB.
The direction cosines $\sin(\theta_{\text{A,L}})$
and $\sin(\theta_{\text{A,N}})$ associated with the 
AoA at the BS are independent random variables 
uniformly distributed into $[0,1]$.
The generic time delay $\tau \in \{\tau_{\text{T},1}, \tau_{\text{T},2}, \tau_{\text{A,L}}, \tau_{\text{A,N}}\}$ of the TU and AU channels is randomly generated according to the
one-sided exponentially decreasing delay power spectrum \cite{Haas}, i.e.,
\be
\tau  = -\tau_{\text{slope}} \, \text{ln} \left[1-u  \left (1-e^{-\Delta_{\text{max}}/\tau_{\text{slope}}}
\right )\right]
\ee
where $\Delta_{\text{max}}=3 \, T_\text{c}$, the slope-time is
$\tau_{\text{slop}} = 2 \, T_\text{c}$, and $u$ is a random variable 
uniformly distributed into $[0,1]$.
Unless otherwise specified, we assume that the constant radial speed of the AU is $v = 8$ m/s,
$f_\text{max}=720$ Hz, and the AoD
$\vartheta_{\text{A,L}}$ and $\vartheta_{\text{A,N}}$ 
are independent random variables 
uniformly distributed into $[0,\pi]$.
The power relationship between the AU and TU at the BS is measured by
the {\em aerial-to-terrestrial ratio (ATR)} defined as
\be
\text{ATR} \eqdef \frac{\sigma_{\text{A,L}}^2+\sigma_{\text{A,N}}^2}
{\sum_{k=1}^{\Ks} \sigma_{g_{\text{T},k}}^2} 
\ee
whereas the signal-to-noise ratio is given by 
\be
\text{SNR} \eqdef \frac{\sigma_{\text{A,L}}^2+\sigma_{\text{A,N}}^2}{\sigma^2_w} \:.
\ee

In each one of the $400$ Monte Carlo runs,  a new set of symbols, noise and channel parameters for both the AU and TU are randomly generated.
Regarding the estimation of the channel parameters of the AU
(see Section~\ref{sec:estimation-AU}), we calculate
the arithmetic mean of the MSEs of the Doppler shifts
$f_{\text{A,L}}$ and $f_{\text{A,N}}$ (normalized by $f_\text{max}^2$),  the time delays
$\tau_{\text{A,L}}$ and $\tau_{\text{A,N}}$ (normalized by $\Delta^2_\text{max}$),
the channel gains  $g_{\text{A,L}}$ 
and $g_{\text{A,N}}$  (normalized by $\sigma^2_\text{A,L}$),
and the direction cosines $\sin(\theta_{\text{A,L}})$
and $\sin(\theta_{\text{A,N}})$.
With reference to the estimation of the TU channel (see Section~\ref{sec:estimation-TU}), we report the estimator output 
\eqref{eq:variance} (normalized by $\sum_{k=1}^{\Ks} \sigma_{g_{\text{T},k}}^2$).
As baseline channel estimation schemes, we report the OMA-based counterparts of the 
proposed estimators, which are implemented by orthogonalizing pilot symbols in the frequency domain, i.e., we 
set $\mathcal{M}_\text{A} \equiv \{0,1,\ldots, 7\}$ and
$\mathcal{M}_\text{T} \equiv \{8,9,\ldots, 15\}$ in each training symbol interval.

\begin{figure}[!t]
\centering
\includegraphics[width=\linewidth]{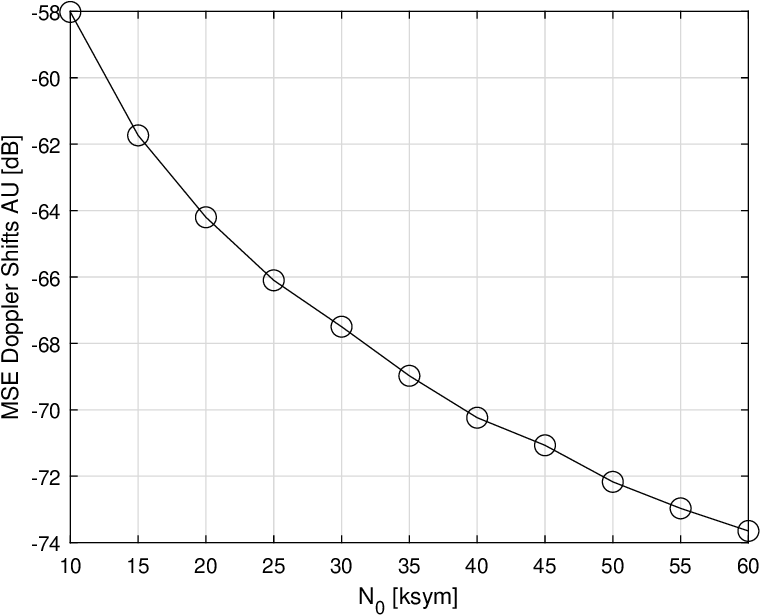}
\caption{MSE of the AU Doppler shifts versus the length $N_0$ of the observation interval, 
($\text{ATR}=0$ dB and $\text{SNR}=14$ dB).}
\label{fig:fig_2}
\end{figure}
\begin{figure}[!t]
\centering
\includegraphics[width=\linewidth]{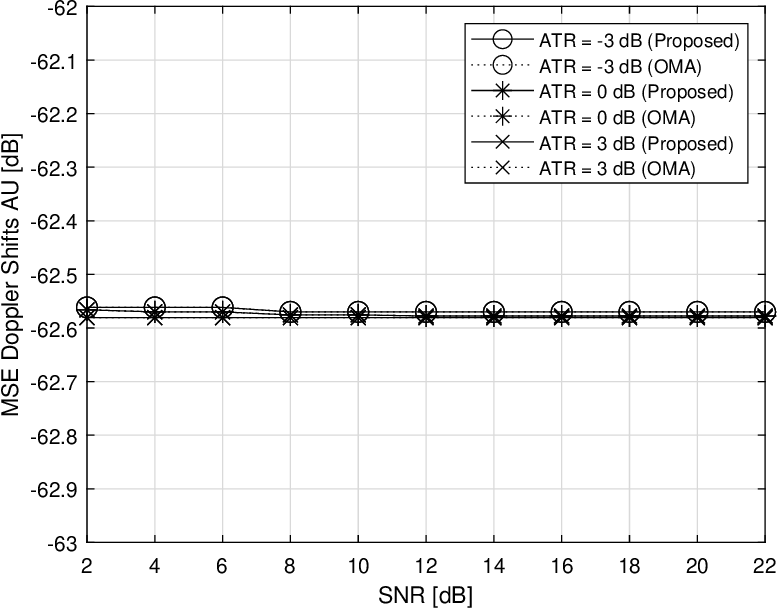}
\caption{MSE of the AU Doppler shifts versus SNR and ATR values.}
\label{fig:fig_3}
\end{figure}
\begin{figure}[!t]
\centering
\includegraphics[width=\linewidth]{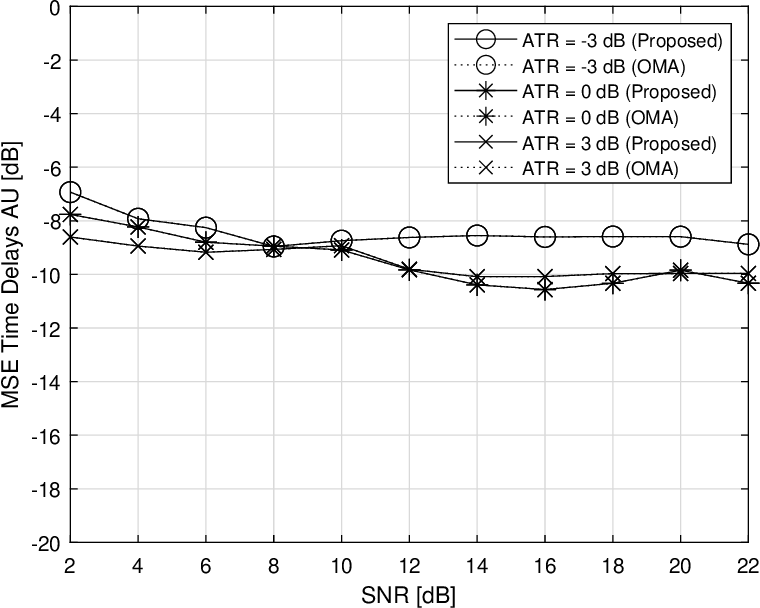}
\caption{MSE of the AU time delays versus SNR and ATR values.}
\label{fig:fig_4}
\end{figure}
\begin{figure}[!t]
\centering
\includegraphics[width=\linewidth]{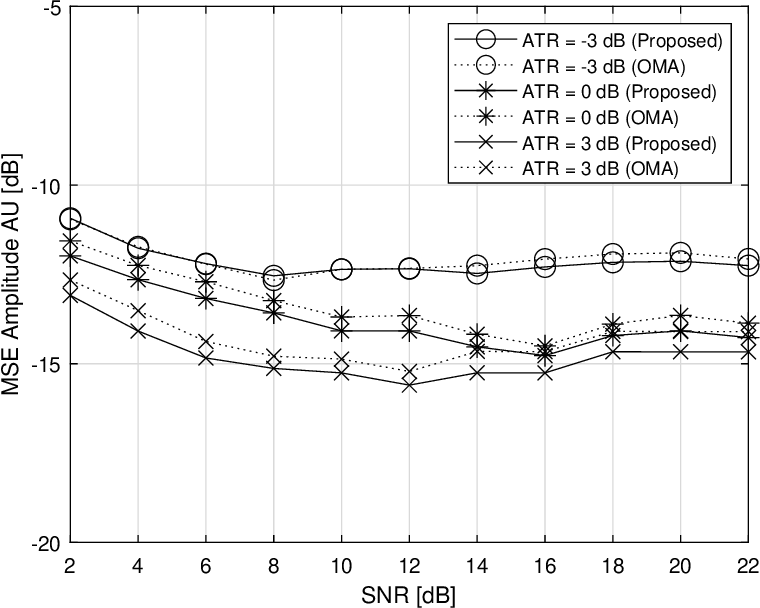}
\caption{MSE of the AU channel gains versus SNR and ATR values.}
\label{fig:fig_5}
\end{figure}
\begin{figure}[!t]
\centering
\includegraphics[width=\linewidth]{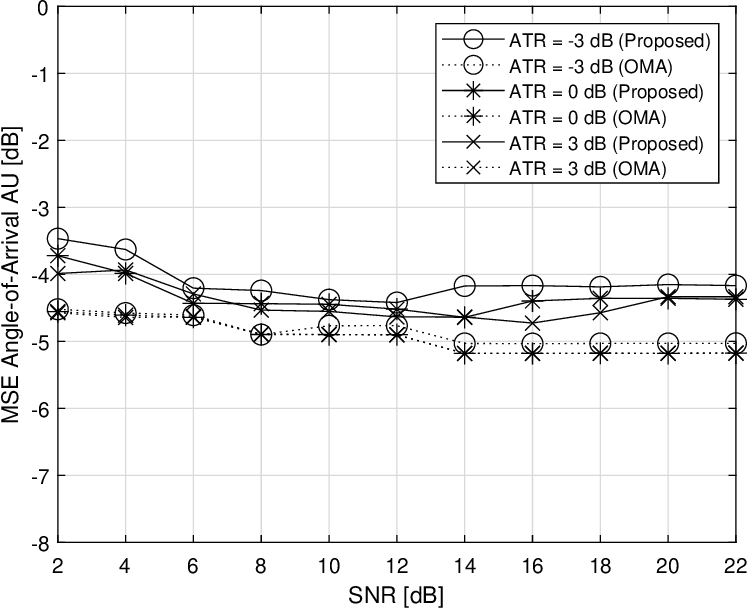}
\caption{MSE of the AU directive cosines versus SNR and ATR values.}
\label{fig:fig_6}
\end{figure}
\begin{figure}[!t]
\centering
\includegraphics[width=1\linewidth]{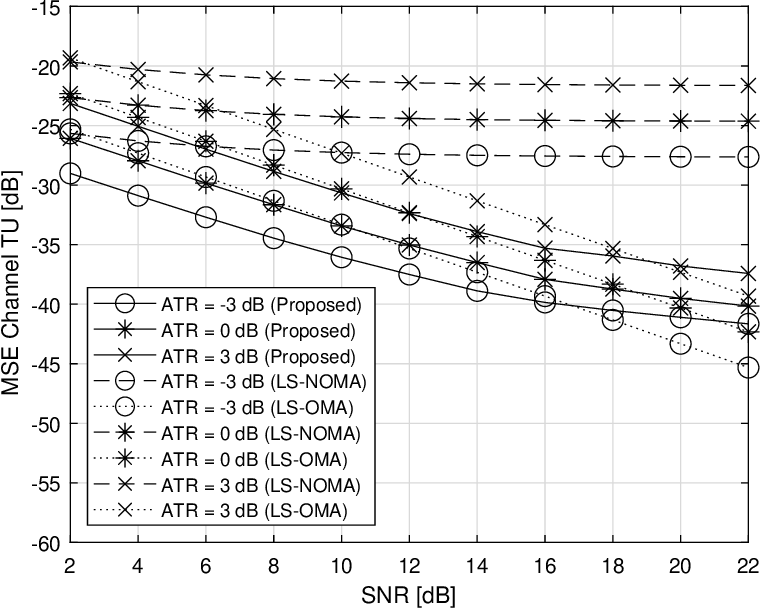}
\caption{MSE of TU channel versus SNR and ATR values
(dashed line for the LS-NOMA estimator, solid line for the BWLU estimator,
and dotted line for the LS-OMA estimator).}
\label{fig:fig_7}
\end{figure}

Figs.~\ref{fig:fig_2}--\ref{fig:fig_6} depict the estimation performance
of the proposed finite-sample 
estimators outlined in Section~\ref{sec:estimation-AU} for
the AU channel parameters (i.e., Doppler shifts, time delays, 
channel gains, and AoAs). 
Generally, it is apparent that such estimators exhibit satisfactory
MSE performance for all the considered SNR and ATR values.

Fig.~\ref{fig:fig_2} reports the MSE
of the Doppler shifts 
as a function of the length $N_0$ of the observation interval 
(in OFDM symbols). As expected, the estimation accuracy 
rapidly improves as $N_0$ increases, thus demonstrating that
the proposed blind estimator of the Doppler shifts is  asymptotically consistent. 
Remarkably, Fig.~\ref{fig:fig_3} shows that the MSE of the Doppler shifts
is independent of both the SNR and ATR, achieving 
the value $-62.57$ dB approximately. 
Such an invariance with respect to 
both the SNR and ATR comes from the fact that 
the blind estimator proposed in 
Subsection~III-A-1 relies on the second-order 
conjugate cyclic statistics of the received
data, for which the contribution of the circular 
wide-sense stationary TU signal and noise
is negligible for sufficiently large (but finite) values of $N_0$. 

We remember that, as the Doppler shifts,  the time delays are blindly estimated by using 
the conjugate second-order statistics of the ACS random process \eqref{eq:y-IBI}.
In this case, a weak dependence of the MSE on the SNR and ATR values can 
be observed in Fig.~\ref{fig:fig_4}, which is due to the fact that 
the estimates of the time delays in Subsection~III-A-2
also feel the aftereffects of the estimation error of the cycle frequency
acquired previously.
Estimation errors due to finite sample-size effects (i.e., when the 
conjugate second-order statistics are estimated 
from a finite number of samples) depend on both
the SNR and ATR in a nonlinear manner \cite{Li.1993}, thus contributing to 
the fluctuations of the curves in Fig.~\ref{fig:fig_4}.
However,  the variability of the MSE of the time delays is confined in a range of about $1$ dB
as a function of the SNR and the MSE curves translate down of nearly $1$ dB  
further to an ATR reduction of  $3$ dB.
It is worth noticing that the performance of the proposed NOMA-based estimators
in Figs.~\ref{fig:fig_3} and \ref{fig:fig_4} is almost indistinguishable from that of the corresponding OMA-based counterparts,
thus corroborating the high robustness of
the proposed Doppler shifts and time delays estimation approaches
against the interference caused by the TU.

The MSE of the channel amplitudes in Fig.~\ref{fig:fig_5} also exhibits a 
weak dependence on the SNR and ATR values. 
However, the 
additional exploitation of
the pilot symbols allows to achieve MSE values smaller than $-10$ dB for 
$\text{SNR}>2$ dB, even when the ATR is as low as $-3$ dB. 
In contrast, the MSE of the AoAs in Fig.~\ref{fig:fig_6} is actually 
independent of the ATR, thus further corroborating the robustness of
the proposed pilot-based estimators developed in Subsection~III-B
against the interference generated by the TU. 
With respect to Figs.~\ref{fig:fig_3} and \ref{fig:fig_4}, 
fluctuations in the MSE curves of channel amplitudes in Fig.~\ref{fig:fig_5} and 
AoAs in Fig.~\ref{fig:fig_6} become even more evident since,
although the corresponding estimation procedures are based on a LS approach
that does not involve estimation of statistics, these estimators suffer of the error propagation 
phenomenon that is inherent of the proposed multi-stage estimation approach 
(see Fig.~\ref{fig:fig_sum_alg}), according to which the existence of estimation 
error in a given stage  will deteriorate the estimation accuracy in the next stage.
Such an error propagation prevents the LS estimates of  channel amplitudes
and AoAs from monotonically decreasing for increasing values of SNR and/or ATR.
The proposed NOMA-based estimators and their  
corresponding OMA-based counterparts in Figs.~\ref{fig:fig_5} and \ref{fig:fig_6} 
exhibit slight performance differences. In particular, 
the performance of OMA-based estimators gets slightly worse than
that of the corresponding NOMA-based counterparts in the case of 
amplitude AU estimation. This behavior is due to the fact that 
amplitude estimation procedures are typically very sensitive to 
the number of available pilot symbols. Indeed, as a consequence of the 
orthogonalization of pilot symbols in the frequency domain, 
OMA-based estimators count on a reduced number of pilot 
symbols, compared to NOMA-based estimation algorithms.

Fig.~\ref{fig:fig_7} depicts the MSE performance of the 
TU channel estimator proposed in Section~\ref{sec:estimation-TU}.
Besides the MSE of the BWLU estimator 
\eqref{eq:BWLU} (solid line), we also report the MSE of the simpler LS-NOMA
estimator \eqref{eq:LS} (dashed line).
As predicted by our analysis, the BWLU estimator largely 
outperforms the LS-NOMA one since it additionally exploits 
the CSI knowledge of the AU. This demonstrates the 
importance of firstly estimate the channel parameters 
of the AU and, then, using the obtained estimates to 
acquire the CSI of the TU.  
As a matter of fact, 
the performance curves of the OMA-based LS estimator (dotted line) are monotonically 
decreasing functions of the SNR, whereas the MSE performance of the corresponding 
NOMA-based counterparts exhibits a floor in the high 
SNR regime, which is due to the residual interference from the 
AU transmission.

\begin{table*}
\caption{MSEs of AU and TU channel parameters for different values of 
the radial speed of the AU [dB] .
}
\label{tab:tab_1}
\centering{}
\begin{tabular}{c|c|c|c|c|c|c|c|c|c|c|c}
\hline
\multirow{2}{*}{\bfseries $v$ [m/s]} & \multicolumn{2}{c|}{\bfseries Doppler Shifts AU} & \multicolumn{2}{c|}{\bfseries Time Delays AU} & \multicolumn{2}{c|}{\bfseries Amplitude AU} & \multicolumn{2}{c|}{\bfseries AoA AU} & \multicolumn{3}{c}{\bfseries Channel TU} \\
\cline{2-12}
& \bfseries Proposed & \bfseries OMA & \bfseries Proposed & \bfseries OMA & \bfseries Proposed & \bfseries OMA & \bfseries Proposed & \bfseries OMA & \bfseries Proposed & \bfseries LS-NOMA & \bfseries LS-OMA \\
\hline
2   &  -62.8159  & -62.8159  & -11.2695  & -11.2695  & -16.2791 & -15.7644  & -2.7585  & -3.9616  & -37.4665  & -24.9258 & -34.1684 \\
4   &  -62.6253  & -62.6253  & -10.8137  & -10.8137  & -15.9419  & -15.7462  & -3.3521  & -4.5738  & -37.1741  & -24.5377 & -34.2129 \\
8   &   -62.5776  & -62.5776  & -10.3907  & -10.3907  & -14.5273  & -14.1755   & -4.6412   & -5.1795  & -36.5062  & -24.5013 & -34.3215  \\
16 & -62.5230  & -62.5230   & -9.7083  &  -9.7083  & -12.8208  & -12.5091  & -6.7672  & -6.0224  & -35.2183  & -24.3936 & -34.8025 \\
\hline
\end{tabular}
\end{table*}

Finally, Tab.~\ref{tab:tab_1} reports the MSE performance of all the estimation algorithms
under comparison for different value of the radial speed $v$ of the AU.
It can be inferred that the radial speed of the AU has a negligible 
effect on the estimates, except for the estimated values of the AoA of
the AU, whose accuracy improves for increasing value of $v$.

\section{Conclusions and directions for future work}
\label{sec:concl}

In this paper, we have considered the 
scenario where a flying AU is paired
with a static TU through power-domain uplink NOMA.
Despite many existing works, we have investigated 
the performance of finite-sample channel estimation procedures. 
We have shown that blind and pilot-aided 
techniques can be suitably combined to 
accurately estimate the AU and TU channel  
that are characterized by different second-order wide-sense 
properties in both the ACS (or Doppler) and modulation domains. 
Specifically, the AU mobility induces ACS features 
in the signal received from sky due to Doppler effects that are not 
present in the signal received from ground,
provided that the TU is static. Moreover, 
the AU can employ a noncircular modu\-lation format, 
as opposed to the TU that transmits circular symbols.
We have demonstrated that Doppler shifts and time delays
of the AU can be estimated blindly with high accuracy 
that is basically independent of the TU and noise powers.
The remaining AU parameters, i.e., channel gains and AoAs,
can be instead estimated through a pilot-assisted LS approach
that exploits the previously acquired knowledge of 
the Doppler shifts and time delays of the AU.
By equipping the BS with the CSI knowledge of the AU, 
the overall channel of the TU can be subsequently acquired  
through a high-precision pilot-based BWLU estimator.
 
The performance of the proposed estimators has been corroborated
through Monte Carlo numerical results. In this respect, an interesting
research subject consists of theoretically investigating the consistency
and asymptotic distribution of such algorithms.
Another appealing research issue is to extend our framework to the case 
in which TUs also communicate with the BS according to a NOMA scheme. 

%
%

\appendix

In BPSK modulation, the symbols
$\{\babpsk[n]\}$ are modeled as a
sequence of  i.i.d. real-valued random
variables, each assuming equiprobable values in $\{\pm 1\}$.
A variation of BPSK is $\pi/2$-BPSK 
modulation scheme \cite{3GPP-1, 3GPP-2, 3GPP-3},
wherein two sets of BPSK constellations are shifted by $\pi/2$, i.e., 
\be
\ba[n]= e^{j \frac{\pi}{4}} \, e^{j \frac{\pi}{2} [n]_2} \, \babpsk[n] \:.
\ee
It can be readily verified that $\Es(|\ba[n]|^2)=1$ and 
$\Es(\ba^2[n])= j \, (-1)^n$.
Consequently, under the standard assumption that the number of subcarriers $M$ is even, one has
$\Es(\bm{s}_\text{A}[n] \, \bm{s}_\text{A}^\herm[n])=\I_{M}$, whereas
$\Es(\bm{s}_\text{A}[n] \, \bm{s}_\text{A}^\trasp[n])=\bm{\Delta}$, with
$\bm{\Delta} \eqdef j \, \diag(1, -1, 1, \ldots, -1) \in \Cset^{M \times M}$.

%
%

\begin{IEEEbiography}[
{\includegraphics[width=1in,height=1.25in,clip,keepaspectratio]{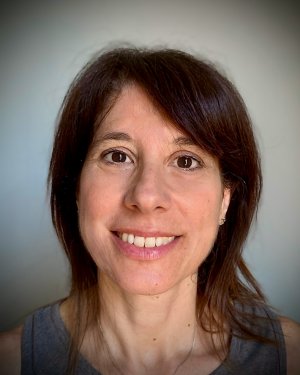}}]
{Donatella Darsena} (Senior Member, IEEE) received the Dr. Eng. degree summa cum laude in telecommunications engineering in 2001, and the Ph.D. degree in electronic and telecommunications engineering in 2005, both from the University of Napoli Federico II, Italy. From 2001 to 2002, she worked as embedded system designer in the Telecommunications, Peripherals and Automotive Group, STMicroelectronics, Milano, Italy. 
In 2005 she joined the Department of Engineering at Parthenope University of Napoli, Italy and worked first as an Assistant Professor and then as an Associate Professor from 2005 to 2022.
She is currently an Associate Professor in the Department of Electrical Engineering and Information Technology of the University of Napoli Federico II, Italy.
Her research interests are in the broad area of signal processing for communications, with current emphasis on reflected-power communications, orthogonal and nonorthogonal multiple access techniques, wireless system optimization, and physical-layer security.
Dr. Darsena has served as a Senior Editor for IEEE ACCESS since 2024, Executive Editor for IEEE COMMUNICATIONS LETTERS since 2023, and Associate Editor for IEEE SIGNAL PROCESSING LETTERS since 2020. She was an Associate Editor of IEEE ACCESS (from 2018 to 2023), of IEEE COMMUNICATIONS LETTERS (from 2016 to 2019), and Senior Area Editor of IEEE COMMUNICATIONS LETTERS (from 2020 to 2023). 
\end{IEEEbiography}

\begin{IEEEbiography}
[{\includegraphics[width=1in,height=1.25in,clip,keepaspectratio]{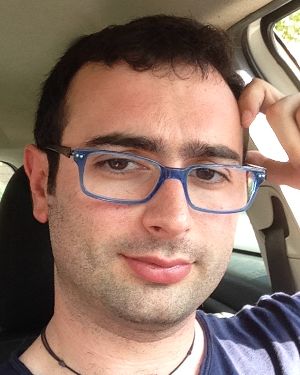}}]
{Ivan Iudice} received the B.S. and M.S. degrees
in telecommunications engineering in 2008 and 2010,
respectively, and the Ph.D. degree
in information technology and electrical engineering in 2017,
all from University of Napoli Federico II, Italy.

Since November 2011,
he has been with the Italian Aerospace Research Centre (CIRA), Capua, Italy.
He first served as part of the Electronics and Communications Laboratory
and he is currently part of the Security Unit.
He is involved in several international projects.
He serves as reviewer for several international journals
and as TPC member for several international conferences.
He is author of several papers on refereed journals and international conferences.
His research activities mainly lie in the area of
signal and array processing for communications,
with current interests focused on physical-layer security,
space-time techniques for cooperative communications systems
and reconfigurable metasurfaces.
\end{IEEEbiography}

\begin{IEEEbiography}[
{\includegraphics[width=1in,height=1.25in,clip,keepaspectratio]{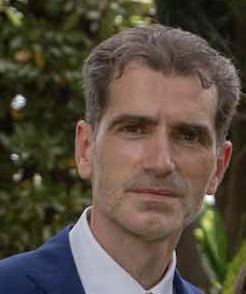}}]
{Francesco Verde}(Senior Member, IEEE)  received the Dr. Eng. degree
\textit{summa cum laude} in electronic engineering
from the Second University of Napoli, Italy, in 1998, and the Ph.D.
degree in information engineering
from the University of Napoli Federico II, in 2002.
Since December 2002, he has been with the University of Napoli Federico II, Italy. He first served as an Assistant Professor of signal theory and mobile communications
and, since December 2011, he has served as an Associate Professor of telecommunications with the Department of Electrical Engineering and Information Technology.
His research activities include reflected-power communications, 
orthogonal/non-orthogonal multiple-access techniques, wireless systems optimization, and 
physical-layer security.

Prof. Verde has been involved in several technical program committees of major IEEE conferences in signal processing and wireless communications.
He has served as Associate Editor for IEEE TRAN\-SACTIONS ON VEHICULAR TECHNOLOGY since 2022.
He was an Associate Editor of the IEEE TRANSACTIONS ON SIGNAL PROCESSING (from 2010 to 2014), IEEE SIGNAL PROCESSING LETTERS (from 2014 to 2018),
IEEE TRANSACTIONS ON COMMUNICATIONS (from 2017 to 2022), and 
Senior Area Editor of the IEEE SIGNAL PROCESSING LETTERS (from 2018 to 2023), 
as well as Guest Editor of the EURASIP Journal on Advances in Signal Processing in 2010 and SENSORS MDPI in 2018-2022.
\end{IEEEbiography}

\end{document}

%% file: arXiv.tex
\usepackage{tikz}

\newcommand\copyrighttext{%
  \footnotesize \textcopyright \the\year{} IEEE. Personal use of this material is permitted. Permission from IEEE must be obtained for all other uses, including reprinting/republishing this material for advertising or promotional purposes, collecting new collected works for resale or redistribution to servers or lists, or reuse of any copyrighted component of this work in other works.}

\newcommand\acceptedtext{%
  \footnotesize This article has been accepted for publication in a future issue of this journal,
  but has not been fully edited. Content may change prior to final publication. \\
  Citation information: DOI 10.1109/OJCOMS.2024.3451308, IEEE Open Journal of Communications Society.}
  
\newcommand\acceptednotice{%
\begin{tikzpicture}[remember picture,overlay]
\node[anchor=north,yshift=2pt,xshift=72pt] at (current page.north) {%
\begin{minipage}{\textwidth}
\center \acceptedtext
\end{minipage}};
\end{tikzpicture}%
}